\documentclass[aps,prx,twocolumn,amsmath,nofootinbib,10pt,floatfix]{revtex4-1}

\usepackage{graphicx}
\usepackage[normalem]{ulem}
\usepackage{xcolor}
\usepackage[utf8]{inputenc}
\usepackage{comment}
\DeclareMathSizes{8}{7}{7}{7}

\usepackage{textgreek}
\usepackage[colorlinks, linkcolor=blue, citecolor=blue]{hyperref}
\usepackage{gensymb}
\usepackage[compat=1.1.0]{tikz-feynman}
\usepackage{tikz}
\tikzset{
pattern size/.store in=\mcSize, 
pattern size = 5pt,
pattern thickness/.store in=\mcThickness, 
pattern thickness = 0.3pt,
pattern radius/.store in=\mcRadius, 
pattern radius = 1pt}

\usepackage{bbold}
\usepackage{mathtools}

\newcommand{\be}{\begin{equation}}
\newcommand{\ee}{\end{equation}}
\newcommand{\beq}{\begin{eqnarray}}
\newcommand{\eeq}{\end{eqnarray}}
\newcommand{\ba}{\[\begin{aligned}}
\newcommand{\ea}{\end{aligned}\]}
\newcommand{\bal}{\begin{aligned}}
\newcommand{\eal}{\end{aligned}}

\renewcommand{\Re}{{\rm Re\,}}

\renewcommand{\vec}[1]{{\bf #1}}
\renewcommand{\epsilon}{\varepsilon}
\renewcommand{\dag}{\dagger}

\renewcommand{\vec}[1]{\boldsymbol{#1}}

\def \q{{\vec{q}}}

\def \r{{\bf {r}}}

\def \ket#1{{\,|\,#1\,\rangle\,}}

\def \tn{\textnormal}

\def \ba{\begin{align*}}
\def \ea{\end{align*}}

\newcounter{indice}

\newcommand{\rcite}[1]{\textcolor{black}{\cite{#1}}}
\makeatletter
\newcommand{\hideappendixsubsectionsintoc}{\let\l@subsection\@gobbletwo}
\makeatother

\begin{document}  
\title{Determinant Quantum-Quantum Monte Carlo: Coherent Auxiliary-Field Sampling}
\author{Xuepeng Wang}\email{xw577@cornell.edu}
\author{Sagnik Banerjee}
\author{Debanjan Chowdhury}
\affiliation{Department of Physics, Cornell University, Ithaca, New York 14853, USA.}
\begin{abstract}
We introduce determinant quantum-quantum Monte Carlo (DQ$^2$MC), a quantum algorithm that lifts the auxiliary-field sampling and averaging at the operational core of determinant quantum Monte Carlo onto a quantum computer.
A determinant oracle synthesizes the DQMC amplitudes directly from a block encoding of the single-particle action matrix via quantum singular value transformations, so that the exponentially many Hubbard-Stratonovich weights are never enumerated, precomputed, or stored. 
Since the fermions are free for fixed auxiliary fields, the construction operates entirely at the single-particle level, requiring $O(\log N_{\tn{st}})$ system qubits and no Jordan-Wigner or Bravyi-Kitaev encoding, where $N_{\tn{st}}$ is the space-time volume.
A full-quantum protocol makes observables interference amplitudes, eliminating the Markov chain and its autocorrelation time altogether; a hybrid quantum-classical protocol retains a constant-size active block of qubits and replaces the Metropolis-Hastings acceptance step with an exact heat-bath draw, so that cluster updates of any size are rejection-free, and passes only classical information between updates, admitting parallel tempering and distributed execution across quantum processors. 
The circuit-depth scales more favorably with spatial volume than classical DQMC, at the price of a post-selection overhead determined exactly by the largest target probability --- polynomial for smooth distributions, exponential for sharply peaked ones.
Finally, the reweighting estimator underlying the fermion sign problem maps exactly onto a quantum weak value, placing the exponential cost of sign-problematic DQMC in precise correspondence with the post-selection overhead of weak-value extraction.
\end{abstract}

\maketitle

\tableofcontents

\section{Introduction}\label{sec:intro}
Understanding the collective behavior of many interacting quantum particles is one of the central challenges of modern physics~\rcite{Feynman1982SimPhys}.
When interactions become comparable to the kinetic energy, perturbative methods fail, and the exponential growth of the many-body Hilbert space imposes a fundamental bottleneck that renders direct computation intractable.  
This difficulty underlies some of the most celebrated open problems in physics, including the mechanism of high-temperature superconductivity~\rcite{Keimer2015Nature,LeeNagaosaWen2006RMP}, the nature of the metallic state near a Mott insulator~\rcite{Stewart2001RMP,FradkinKivelsonTranquada2015RMP,PhillipsHusseyAbbamonte2022Science}, and the quark-gluon phase diagram at finite baryon density~\rcite{AlfordRajagopalWilczek1998CSC,StephanovRajagopalShuryak1998CEP}, just to name a few.
For fermions, the workhorse ``classical" algorithms face a sharp divide: in sign-problem-free regimes, unbiased finite-temperature simulation is possible but expensive at scale, while in sign-problematic regimes the cost grows exponentially with system size and inverse
temperature~\rcite{LohGubernatis1990sign,TroyerWiese2005}.  
In this work, we show that the operational core of the leading classical algorithm for the first regime, namely the sampling and averaging over auxiliary fields,  can be lifted directly onto a quantum device, with the second regime recast, in a precise operational sense, as a problem in quantum measurement theory.
 
Determinant quantum Monte Carlo (DQMC) is the canonical unbiased method for simulating finite-temperature properties of strongly correlated lattice fermions in sign-problem-free
regimes~\rcite{Hirsch1983discrete,Sorella1989,AssaadEvertz2008}. 
Its status is earned by elimination: exact diagonalization is confined to small clusters by the exponential Hilbert space; tensor-network methods~\rcite{White1992DMRG,Schollwoeck2011RMP,VerstraeteMurgCirac2008PEPS} are constrained by entanglement-area laws in two and three dimensions; dynamical mean-field theory and its cluster extensions~\rcite{Georges1996DMFT,Maier2005CDMFT} treat inter-site fluctuations approximately; and auxiliary-field and diagrammatic Monte Carlo schemes~\rcite{ZhangKrakauer2003,ProkofevSvistunov1998DiagMC} are ultimately limited by the fermion sign problem~\rcite{LohGubernatis1990sign,TroyerWiese2005}. 
By decoupling the interaction via a Hubbard-Stratonovich (HS) transformation, DQMC recasts the interacting partition function as a statistical average over auxiliary-field configurations, each of which is a free-fermion problem weighted by a fermion determinant.  The method has become indispensable for studying the interplay of high-temperature superconductivity and quantum criticality~\rcite{annrevQMC}, density-wave order~\rcite{TDstripes}, and non-Fermi-liquid metals~\rcite{TDsm}, in platforms ranging from transition-metal oxides to moir\'e quantum
materials, where interaction strength, bandwidth, filling, and lattice geometry can now be engineered with remarkable
precision~\rcite{Cao2018CI,Cao2018SC,AndreiMacDonald2020NatMater,KennesEtAl2021NatPhys}.
Closely analogous auxiliary-field structures arise in lattice quantum chromodynamics, where
confinement~\rcite{Wilson1974LatticeQCD,Creutz1980LatticeMC}, chiral-symmetry breaking~\rcite{GrossWilczek1973,Shuryak1980Plasma}, and the finite-density phase
diagram~\rcite{AlfordRajagopalWilczek1998CSC,StephanovRajagopalShuryak1998CEP,BraunMunzingerStachel2007Nature} pose the same computational challenges for simulations of fermions
coupled to dynamical gauge
fields~\rcite{Fucito1981FermionMC,Karsch2002LatticeReview,FodorKatz2002Reweight,BorsanyiEtAl2010ChiralTc,HotQCDEoS2014,deForcrand2010QCDsign}.
Despite this broad utility, DQMC faces two intrinsic bottlenecks: the cubic cost of evaluating each fermion determinant, and the autocorrelation problem of Markov-chain sampling, both of which become severe in precisely the regimes of greatest physical interest.
 
Quantum simulation, the use of one controllable quantum system to imitate the dynamics of another, offers a fundamentally different path to these problems.  
Feynman's observation that classical computers face an exponential bottleneck in simulating quantum mechanics~\rcite{Feynman1982SimPhys} and Lloyd's proof that a universal quantum computer can efficiently simulate any local Hamiltonian via Trotterized evolution~\rcite{Lloyd1996Sim} placed this vision on rigorous footing, and the field has since matured across digital and analog platforms~\rcite{GeorgescuAshhabNori2014,Altman2021PRX,DaleyEtAl2022Nature}.
Algorithms now exist for ground-state and spectral properties via Trotterized real- and imaginary-time evolution~\rcite{AbramsLloyd1997,Motta2020QITE,McArdle2019VITE}, variational quantum eigensolvers~\rcite{Peruzzo2014VQE,Kandala2017HEVQE,GoogleAI2020HartreeFock}, and qubitization-based methods achieving near-optimal query
complexity~\rcite{LowChuang2017QSP,LowChuang2019qubitization,GilyenSuLowWiebe2019QSVT,Babbush2018encoding,AspuruGuzik2005}.
Quantum counterparts of classical many-body methods have also emerged: sequential and variational state-preparation protocols for matrix-product and PEPS
states~\rcite{Schon2005SequentialMPS,SchwarzTemmeVerstraete2012PEPS,FossFeig2021HolographicMPS},
hybrid quantum-classical dynamical mean-field
theory~\rcite{BauerWeckerMillisHastings2016QDMFT,Kreula2016QDMFT}, and quantum-assisted projector Monte Carlo~\rcite{Huggins2022NatureQCQMC}.
For finite-temperature properties, i.e. the regime most natural to DQMC, proposals include quantum Metropolis sampling~\rcite{Temme2011Metropolis}, Gibbs-state preparation via quantum walks~\rcite{PoulinWocjan2009Gibbs} and Lindbladian dynamics~\rcite{ChenKastoryanoBrandaoGilyen2023ThermalPrep,ChenKastoryanoGilyen2024GibbsSampler,ChenHuangPreskillZhou2024NaturePhys,LinTong2020NearOptimal,DongLinTong2022Eigenvalue,DingLin2024Lindbladian}, and related thermal-state constructions~\rcite{ChowdhurySomma2017Gibbs}.  All of these approaches, however, operate on the many-body Hamiltonian in second quantization; none directly addresses the possibility of auxiliary-field sampling and averaging on quantum processors that constitutes the operational core of DQMC. 
In parallel developments, quantum algorithms have been implemented to promote classically disordered variables to qubits~\rcite{Gyawali2026DisorderFree}, which conceptually motivates the investigation of other applications of quantum promotion of classical variables. 

\begin{figure}[htb!]
\centering
\includegraphics[width=8cm]{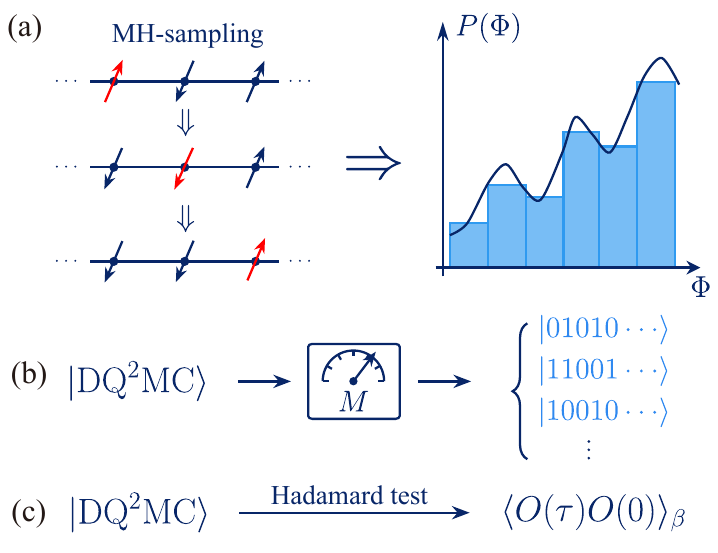}
\caption{\label{Fig::schematics} {\bf Schematic representation of the key ideas.} (a) Classical DQMC generates a Markov chain of Hubbard–Stratonovich configurations through local single-field updates (in red); after equilibration, the histogram of retained configurations approaches the target distribution. (b) $\mathrm{DQ^2MC}$ prepares a coherent superposition of all configurations. Computational-basis measurements directly sample configurations from the superposition. (c) A Hadamard test coherently evaluates the ensemble-averaged observable. In (b) and (c), part of the auxiliary-field control qubits can be downgraded to classical control bits optionally; see the hybrid protocol of Sec.~\ref{sec:compat} and its schematic representation in Fig.~\ref{Fig::fig3}.}
\end{figure}
 
In this work, we introduce Determinant Quantum-Quantum Monte Carlo (DQ$^2$MC), a quantum algorithm that lifts the HS sampling and averaging of classical DQMC entirely onto a quantum computer.
A schematic representation of the key protocols appear in Fig.~\ref{Fig::schematics}.
The central idea is to \textit{coherently encode} the DQMC probability distribution; rather than drawing classical samples from the HS measure (Fig.~\ref{Fig::schematics}a), we prepare a quantum state $|\mathrm{DQ^2MC}\rangle$ whose amplitudes in the auxiliary-field basis are the ``square-roots" of the DQMC weights (Fig.~\ref{Fig::schematics}b), so that physical observables are extracted by quantum measurement in place of stochastic averaging (Fig.~\ref{Fig::schematics}c).  
The sign-problem-free structure is precisely what makes this encoding natural: the anti-unitary symmetry that guarantees positive weights expresses each weight as a squared modulus, $p_{\Phi}\propto|\det M^{+}_{\Phi}|^{2}$, so that the required \textit{amplitude} $\sqrt{p_{\Phi}}$ is simply the single-flavor determinant magnitude; the Born rule performs the flavor doubling. 
Crucially, the amplitudes are \textit{synthesized directly from the microscopic Hamiltonian}, without enumerating, precomputing, or storing the exponentially many HS weights that a generic quantum probability loader~\rcite{GroverRudolph2002,GiovannettiLloydMaccone2008QRAM,KerenidisPrakash2017} would require. 
This exploits the fact that at fixed HS configuration the fermion problem is free.  
The construction therefore operates entirely at the level of the single-particle action matrix: it uses a logarithmic number of system qubits, avoids Jordan-Wigner and Bravyi-Kitaev string operators~\rcite{JordanWigner1928,BravyiKitaev2002Fermion} entirely, and achieves a circuit depth that scales \emph{sub-cubically} in the spatial volume. 
The algorithm thus achieves a more favorable circuit-depth scaling than classical DQMC, at the price of a polynomial overhead in imaginary time and a post-selection overhead set by the condition number of the DQMC distribution itself: polynomial when the distribution is smooth, but potentially exponential in the space-time volume when it is sharply peaked.  
We characterize this overhead precisely, isolate its origin in the probability-loading step, and discuss possible strategies to mitigate it in the outlook.
 
We develop two complementary protocols in this work.  
The \textit{fully-quantum} protocol maintains the entire auxiliary-field register as qubits and coherently averages observables across the full HS distribution via a Hadamard-test estimator, eliminating the autocorrelation time of classical Markov-chain sampling altogether.  
The \textit{hybrid quantum-classical} protocol demotes the bulk of the auxiliary register to classical bits, retaining only a constant-size active block of qubits: preparation and measurement of the active block realizes an exact heat-bath update of the corresponding auxiliary fields, so that cluster updates of any size proceed with no Metropolis-Hastings acceptance step and no additional oracle cost per update relative to single-site moves.  
Observable estimation proceeds by classical averaging over the sampled configurations, with the per-sample observable evaluated coherently on the active block; the qubit overhead can thereby scale as $O(\log(N_{\tn{st}}))$ at optimum without sacrificing the advantage in observable estimation, where $N_{\tn{st}}$ denotes the space-time volume.  
Because only classical information passes between successive updates, the hybrid protocol accommodates classical sampling-acceleration techniques such as parallel tempering and lifted Gibbs samplers without modification, and fits naturally into a distributed quantum computation framework with only classical inter-node communication.
As a benchmark, we demonstrate quantitative agreement between classical DQMC and classical emulation of DQ$^2$MC for the half-filled square-lattice Hubbard model (albeit for small sizes) in an appendix.

Finally, we have also established a precise conceptual correspondence between the fermion sign problem and quantum measurement theory.  
The standard reweighting estimator underlying the sign problem, in which a phase-quenched ensemble is reweighted by the fermion sign to recover the true average, maps \textit{exactly} onto a quantum weak-value protocol~\rcite{AharonovAlbertVaidman1988,DresselRMP2014}, with pre-selection on the phase-quenched state and post-selection on the full, sign-carrying state.  
This identification is not merely formal: the exponential shot overhead intrinsic to weak-value extraction is in precise correspondence with the exponential cost of classical reweighting.  
The map does not remove the sign problem: standard weak-value extraction inherits the same exponential overhead as classical reweighting.
 
The remainder of this paper is organized as follows. 
In Section~\ref{sec:overview-dqmc} we provide a brief review of DQMC and set up our notation. Section~\ref{sec:qalg} presents the full-quantum protocol: state preparation (Sec.~\ref{sec:qalg_state_prep}), coherent observable measurement (Sec.~\ref{sec:qalg_obs}), postselection and hyperparameter scaling (Sec.~\ref{sec:loading}), and resource scaling (Sec.~\ref{sec:qalg_summary}); the detailed determinant-oracle and gate construction is described in Appendix~\ref{app:state_prep_oracle}. 
In Section~\ref{sec:compat} we develop the hybrid quantum-classical protocol, comprising the cluster-Gibbs sampler (Sec.~\ref{subsec:gibbs}) and the corresponding observable
estimator (Sec.~\ref{subsec:hybrid_obs}).
The benchmark setup, numerical results, and preconditioning procedure are discussed in Appendix~\ref{app:hubbard_benchmarks}.
Section~\ref{sec:related} positions DQ$^{2}$MC against prior quantum approaches to Monte Carlo. 
Section~\ref{sec:extensions} develops extensions and generalizations: the zero-temperature projective protocol PQ$^{2}$MC (Sec.~\ref{sec:pq2mc}), the weak-value formulation of the fermion sign problem (Sec.~\ref{sec:sign}), higher-order Gauss-Hermite quadratures (Sec.~\ref{sec:higher_GH}), and a discussion of circumstances in which the auxiliary-field route is preferable to direct simulation of the Hamiltonian (Sec.~\ref{sec:why_aux}). 
We conclude in Sec.~\ref{sec:conclusion} with an outlook, and provide additional technical details in supporting appendices.

\section{Overview of DQMC}\label{sec:overview-dqmc}

Before presenting our quantum algorithm, we briefly review the key ingredients of DQMC, which is a numerically exact auxiliary-field method for interacting fermion systems at finite temperature. 
The method is exact up to a controlled Trotter discretization error that is extrapolated away in practice~\cite{BSS1981,Hirsch1983discrete,Sorella1989,LohGubernatis1990sign,AssaadEvertz2008}. The basic idea is to rewrite the partition function of interacting fermions as a statistical sum over auxiliary-field configurations, in which the system is effectively non-interacting for each configuration. 
Consider a generic interacting fermion Hamiltonian of the form
\begin{subequations}
\beq
\hat{H}&=&\hat{T}+\hat{H}_{\mathrm{int}},\\
\hat{H}_{\mathrm{int}} &=& g \sum_{r} \bigg[ \hat{V}_{r} - \alpha \bigg]^2,
\eeq
\end{subequations}
where $\hat{T}$ is quadratic in the fermion operators and $\hat{H}_{\mathrm{int}}$ is the interaction term ($g\equiv ~\mathrm{interaction~ strength}$) written as a perfect square of a fermion bilinear $\hat{V}_{r}$ at spatial site $r$ with offset $\alpha$. Starting from the partition function of the interacting system
\begin{equation}
Z=\mathrm{Tr}\, e^{-\beta \hat{H}},
\end{equation}
we discretize imaginary time as $\beta=L_{\mathrm{T}}\Delta\tau$ and perform a Trotter decomposition~\cite{Trotter1959,Suzuki1976},
\begin{equation}
e^{-\beta \hat{H}}\approx \prod_{\tau=1}^{L_{\mathrm{T}}}e^{-\Delta\tau \hat{H}_{\mathrm{int}}}e^{-\Delta\tau \hat{T}},
\end{equation}
which carries an $O(\Delta\tau^{2})$ error per time slice, i.e. an $O(\Delta\tau)$ error at fixed $\beta$.
After the HS decoupling~\cite{Stratonovich1957HS,Hubbard1959HS}, the contribution from the quartic interaction at the time slice $\tau$ is rewritten as a sum over auxiliary fields that are linearly coupled to fermion bilinears,
\begin{subequations}
\beq
e^{-\Delta\tau \hat H_{\mathrm{int}}}
&=& \prod_{r}\sum_{\{\Phi_{r,\tau}\}} \gamma[\Phi_{r,\tau}]
e^{V[\Phi_{r,\tau}]},\label{eq::hs}
\\
V[\Phi_{r,\tau}] &\equiv& \sqrt{-g\Delta\tau} ~ \Phi_{r,\tau} \bigg[\hat{V}_{r} - \alpha \bigg],
\eeq
\end{subequations}
where $\Phi_{r,\tau}$ is the auxiliary field defined on spatial site $r$ and imaginary-time slice $\tau$, and $\gamma[\Phi_{r,\tau}]$ is the weight assigned by the Gaussian-quadrature discretization of the underlying Gaussian integral~\cite{Hirsch1983discrete,MotomeImada1997}. For the spin-$1/2$ Hubbard interaction, the two-node (Ising) discrete transformation~\cite{Hirsch1983discrete} is exact; for generic perfect-square interactions, an $n$-node Gauss-Hermite quadrature controls the residual error order (see Sec.~\ref{sec:higher_GH}). 
For a fixed HS configuration $\Phi\equiv\{\Phi_{r,\tau}\}$, the fermions become quadratic and are described by the single-slice propagator,
\begin{equation}
B_{\tau}\equiv e^{\sum_{r} V[\Phi_{r,\tau}]}\,e^{-\Delta\tau \hat{T}}.
\end{equation}
The fermions can then be integrated out exactly via the coherent-state path integral, yielding a sum over the space-time auxiliary-field configurations for the partition function,
\begin{subequations}\label{eq::Z_dqmc}
\beq
Z&=&\sum_{\{\Phi\}}\Gamma[\Phi]\,\det M_{\Phi}, \qquad \Gamma[\Phi]\equiv\prod_{r,\tau}\gamma[\Phi_{r,\tau}], ~~~~~\\
\det M_{\Phi}&=&\det\left(1+B_{L_{\mathrm{T}}}\cdots B_{2}B_{1}\right),
\eeq
\end{subequations}
where the block matrix $M_{\Phi}$ is referred to as the fermion action matrix, defined as
\begin{equation}\label{eq::def_action_matrix}
M_{\Phi}=\left(\begin{array}{ccccc}
1 & 0 & \cdot & 0 & B_{L_{\mathrm{T}}} \\
-B_1 & 1 & 0 & \cdot & 0 \\
0 & -B_2 & 1 & \cdot & 0 \\
\cdot & \cdot & \cdot & \cdot & \cdot \\
0 & \cdot & \cdot & -B_{L_{\mathrm{T}}-1} & 1
\end{array}\right).
\end{equation}
For the Ising-type decoupling used throughout Secs.~\ref{sec:qalg}--\ref{sec:compat}, $\gamma[\Phi_{r,\tau}]=1/2$ is configuration-independent, so $\Gamma[\Phi]$ drops out of all weight ratios and we suppress it below; it becomes configuration-dependent, and must be retained, for the higher-order quadratures of Sec.~\ref{sec:higher_GH}.
Eq.~\eqref{eq::Z_dqmc} shows that the original interacting fermion problem is mapped onto a free-fermion problem in a fluctuating, imaginary-time-dependent auxiliary field, where each auxiliary-field configuration is weighted by a fermion determinant. 
$M_{\Phi}$ is block-bidiagonal (up to the corner block enforcing anti-periodic boundary conditions) and depends on the auxiliary fields only through the one-body vertices inside each $B_{\tau}$, i.e. precisely the structure that the construction in Appendix~\ref{app:state_prep_oracle} block-encodes.
The sum over auxiliary-field configurations in Eq.~\eqref{eq::Z_dqmc} is performed by the Monte Carlo method.
In sign-problem-free settings (see e.g. symmetry-based sufficient conditions such as time-reversal-invariant flavor pairing and Majorana positivity~\cite{WuZhang2005SignFree,LiJiangYao2015Majorana,Wei2016MajoranaPositivity}) one has $\det M_{\Phi}\ge 0$, which therefore defines a normalized probability distribution
\begin{equation}\label{eq::def_prob}
p_{\Phi}=\frac{\det M_{\Phi}}{\sum_{\{\Phi'\}}\det M_{\Phi'}},
\end{equation}
which is statistically sampled in classical DQMC; outside these settings, sampling requires reweighting at exponential cost~\cite{LohGubernatis1990sign,TroyerWiese2005} (see Sec.~\ref{sec:sign}). 
Here $\det M_{\Phi}$ denotes the determinant over all fermion flavors; for the symmetry class considered in Sec.~\ref{sec:qalg}, it factorizes as $\det M_{\Phi}=|\det M^{+}_{\Phi}|^{2}$ into conjugate single-flavor blocks, and from Sec.~\ref{sec:qalg} onward $M_{\Phi}$ denotes the single-flavor block. 
The Monte Carlo sampling over auxiliary-field configurations is typically through local updates of the HS fields accepted according to the corresponding determinant ratios using the Metropolis-Hastings method~\cite{Metropolis1953,Hastings1970}.
 
Physical observables are obtained by averaging over this ensemble. 
The thermal expectation value of an operator $\hat{A}$ can be calculated as,
\begin{equation}
\langle \hat{A}\rangle=\sum_{\{\Phi\}}p_{\Phi}A_{\Phi}.
\end{equation}
Typically in DQMC simulations, $\hat{A}$ is a correlation function of local fermion bilinears $O_\r$. 
In two spatial dimensions, a representative example is the imaginary-time (Matsubara) susceptibility $\chi_{O}(\q,i\Omega_n)$ defined as
\begin{equation}
\label{eq:def_corr}
\chi_{O}(\q,i\Omega_n)=\frac{1}{N_{\tn{st}}}\left\langle\sum_{\r\tau;\r'\tau'}O_{\r}(\tau)O_{\r'}(\tau')e^{i\q\cdot(\r-\r')+i\Omega_n(\tau-\tau')}\right\rangle,
\end{equation}
with bosonic Matsubara frequency $\Omega_n=2\pi n/\beta$ and $N_{\tn{st}}$ the space-time volume. 
Real-frequency spectra such as the dynamical structure factor are related to Eq.~\eqref{eq:def_corr} only through numerical analytic continuation~\cite{JarrellGubernatis1996MaxEnt,Sandvik1998SAC}, which lies outside the scope of this work.
Since the system is non-interacting at fixed auxiliary-field configuration $\Phi$, Wick's theorem reduces the correlators to products of Green's functions.  
In particular, the time-ordered single-particle Green's function is given by the inverse of the fermion action matrix, $G_{\Phi}=M_{\Phi}^{-1}$; on the quantum side, this inverse is obtained via quantum singular value transformation (QSVT) \cite{GilyenSuLowWiebe2019QSVT} (Sec.~\ref{sec:qalg_obs}).
 
Let us end this overview with the classical cost against which our quantum protocols will be compared. 
A local update changes the action matrix by a low-rank perturbation, so the equal-time Green's function can be updated at $O(N_s^{2})$ cost per accepted move~\cite{BSS1981,Hirsch1985Hubbard,White1989Hubbard}; a full sweep over the $N_{\tn{st}}=L_{\tn{T}}N_s$ space-time auxiliary fields therefore costs $O(L_{\tn{T}} N_s^{3})$, and at low temperatures the products of $B_{\tau}$ matrices additionally require numerical stabilization~\cite{White1989Hubbard,AssaadEvertz2008,ALF2022}. Local updates are swept through the space-time lattice, and the Green's function is updated after each accepted move; the sampling is repeated until a convergence threshold is reached, and the final observable estimate is obtained by averaging over the sampled configurations, with a statistical error controlled by the autocorrelation time of the Markov chain. The cubic cost per configuration generated and the autocorrelation time are the two classical bottlenecks that the quantum algorithm of Sec.~\ref{sec:qalg} is designed to address.
\section{Fully Quantum Algorithm}\label{sec:qalg}

We now describe the full-quantum protocol for DQMC. The key is to promote each auxiliary field $\Phi_{r,\tau}$ to a quantum register and to coherently prepare a pure state, denoted by $\ket{\mathrm{DQ}^{2}\mathrm{MC}}$, whose Born probability in the auxiliary-field basis reproduces the DQMC distribution in Eq.~\eqref{eq::def_prob}.
Unless stated otherwise, we write the algorithmic formulas for one orbital and one independent binary HS channel per space-time point; for generic models, matrix dimensions and auxiliary-field counts acquire O(1) factors $n_{\tn{orb}}$ and $n_{\tn{vert}}$, denoting the number of orbitals and vertices per site respectively.
We focus on sign-problem-free systems satisfying the following assumptions:
\begin{enumerate}
    \item The fermions have two flavors $\sigma=\pm$ related by an anti-unitary symmetry $\mathcal{T}$ that acts configuration-by-configuration on the auxiliary-field problem; this is the standard symmetry-based sufficient condition for sign-problem-free DQMC~\cite{WuZhang2005SignFree}.
    \item After the HS decomposition defined in Eq.~\eqref{eq::hs}, the auxiliary-field vertex $e^{V[\Phi_{r,\tau}]}$ is unitary. 
\end{enumerate}
The first assumption gives a fermion action matrix that is block-diagonal in flavor space, where the two blocks are related by $\mathcal{T}$ for every fixed configuration $\Phi$. For each flavor $\sigma$, the matrix $M_{\Phi}^{\sigma}$ has the form in Eq.~\eqref{eq::def_action_matrix}; the anti-unitary symmetry implies $\det M_{\Phi}^{-}=(\det M_{\Phi}^{+})^{*}$, so that the full determinant is $|\det M_{\Phi}^{+}|^{2}$, as anticipated in Sec.~\ref{sec:overview-dqmc}. We therefore work with a single flavor sector, denoted simply by $M_{\Phi}$ below. As emphasized in Sec.~\ref{sec:intro}, this squared-modulus structure is precisely what makes the amplitude encoding natural: the target amplitude $\sqrt{p_{\Phi}}$ is the single-flavor determinant magnitude, and the Born rule performs the flavor doubling.
The second assumption ensures that the HS-dependent one-body vertices can be implemented as controlled unitaries; if this condition is relaxed, these vertices as non-unitary matrices must themselves be block encoded into unitary, leading to an additional postselection overhead. 
 
\begin{figure*}[htb]
\includegraphics[width=18cm]{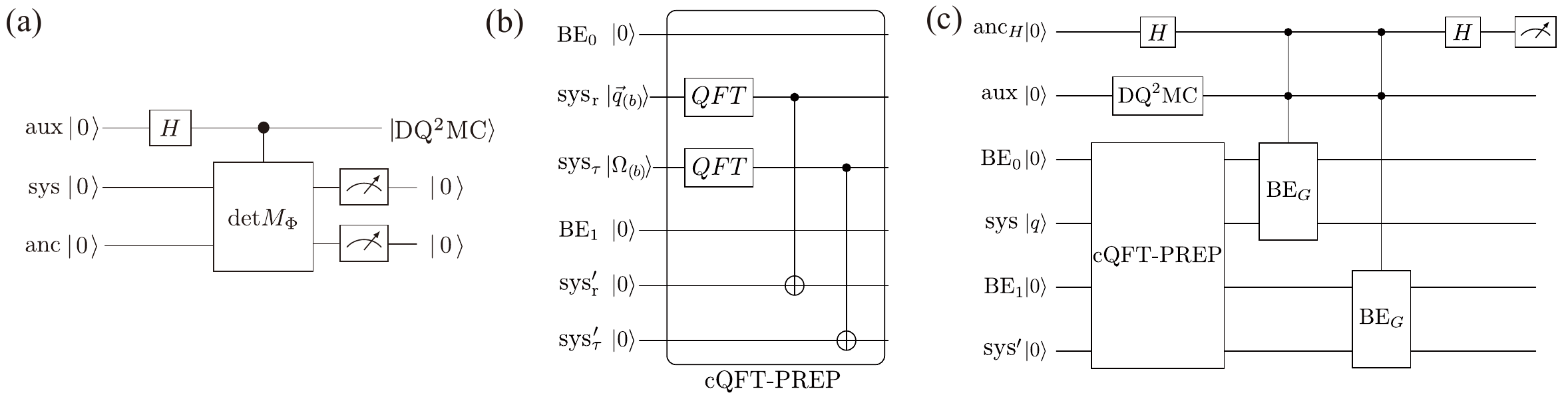}
\caption{\label{Fig::fig_fullq_gates_main} {\bf Overview of the full quantum protocol.} (a) State-preparation circuit: the determinant oracle $\mathcal{O}_{\det}$ (boxed gate $\det M_{\Phi}$) is applied to the Hadamard-prepared auxiliary-field register $\mathtt{aux}$; postselecting all work registers (drawn separately as $\mathtt{sys}$ and $\mathtt{anc}$) on $|0\rangle$ leaves $\ket{\mathrm{DQ}^{2}\mathrm{MC}}$ of Eq.~\eqref{eq:qalg_target} on $\mathtt{aux}$.
(b) Gate realization of cQFT-PREP defined in Eq.~\eqref{eq:qalg_cqft}, where QFT stands for quantum Fourier transform. The registers $\mathtt{sys}_r$ and $\mathtt{sys}_\tau$ denote the qubits for spatial and temporal action-matrix entries, respectively. The register $\mathtt{sys}'$ is a copy of $\mathtt{sys}$. (c) Observable-measurement circuit: a Hadamard test on $\mathtt{anc}_H$ controls two Green's-function block encodings $\mathrm{BE}_G$ acting on a system-copy pair $\mathtt{sys},\mathtt{sys}'$. The system copies are prepared in the coherent momentum-pair state by the cQFT-PREP block, with $\mathtt{aux}$ in $\ket{\mathrm{DQ}^{2}\mathrm{MC}}$. }
\end{figure*}

The construction follows the same logic as the classical method: one assigns a determinant weight to each HS configuration and then evaluates observables by averaging free-fermion correlation functions over the auxiliary-field distribution. The essential difference is that the sum over HS configurations is carried out coherently.
Fig.~\ref{Fig::fig_fullq_gates_main} summarizes the protocols for state preparation and coherent observable measurement. We first define the target state and determinant-oracle interface (Sec.~\ref{sec:qalg_state_prep}; Appendix~\ref{app:state_prep_oracle}), then describe coherent observable measurement (Sec.~\ref{sec:qalg_obs}, Fig.~\ref{Fig::fig_fullq_gates_main}), analyze postselection and hyperparameter scaling (Sec.~\ref{sec:loading}), and summarize resources (Sec.~\ref{sec:qalg_summary}).

\subsection{State Preparation}\label{sec:qalg_state_prep}

Before describing the construction in detail, it is useful to state its logic briefly. The target operation of multiplying each auxiliary-field branch by its determinant weight is non-unitary. The standard remedy is block encoding, in which the desired operator is embedded as a sub-block of a larger unitary and realized probabilistically by post-selecting ancilla registers.
The determinant itself is a product over all $N_{\tn{st}}$ singular values and can therefore be synthesized the same way one computes it classically: linearize via the logarithm, $\log|\det M_{\Phi}|=\operatorname{Tr}\log\sqrt{M_{\Phi}^{\dagger}M_{\Phi}}$,
then exponentiate the resulting scalar.

We formulate the state-preparation problem by designing a determinant oracle that synthesizes the DQMC weight. As described above, the full fermion determinant factorizes into two symmetry-related flavor sectors. We denote by $M_{\Phi}$ the action matrix for one flavor sector and suppress the flavor label. The DQMC probability distribution can then be written as
\begin{equation}
p_{\Phi}
=
\frac{|\det M_{\Phi}|^2}{\sum_{\Phi'}|\det M_{\Phi'}|^2},
\label{eq:qalg_pphi}
\end{equation}
where the square accounts for the two conjugate flavor sectors. The target state for the auxiliary-field register (denoted as $\mathtt{aux}$ in Fig.~\ref{Fig::fig_fullq_gates_main}a) is
\begin{equation}
|\mathrm{DQ}^{2}\mathrm{MC}\rangle
=
\sum_{\{\Phi\}}\sqrt{p_{\Phi}}\,|\Phi\rangle_{\mathtt{aux}}
\;\propto\;
\sum_{\{\Phi\}}|\det M_{\Phi}|\,|\Phi\rangle_{\mathtt{aux}}.
\label{eq:qalg_target}
\end{equation}
The desired amplitudes can be encoded in a diagonal determinant operator, defined as
\begin{equation}
D_{\det}\equiv \sum_{\{\Phi\}}|\det M_{\Phi}|\,|\Phi\rangle\langle\Phi|,
\label{eq::det_gate_def}
\end{equation}
which is not unitary in general.

We seek an oracle $\mathcal{O}_{\det}$, as a block encoding (BE) of the diagonal operator defined in Eq.~\eqref{eq::det_gate_def}, such that pre- and post-selecting the block-encoding ancilla on $|0\rangle_{\mathtt{anc}}$ realizes the desired operator $D_{\det}$ on the auxiliary-field register.
As shown in Fig.~\ref{Fig::fig_fullq_gates_main}(a), the BE ancilla consists of two parts: a system register $\mathtt{sys}$ used to load the fermion action matrix $M_{\Phi}$ controlled by the auxiliary-field register, and additional ancilla qubits $\mathtt{anc}$ for realizing the determinant oracle. The detailed composition of $\mathtt{anc}$ will be explained later. With $\mathbb{1}_{\Phi}$ denoting the identity operator in auxiliary-field register and $\mathtt{sa}\equiv\{\mathtt{sys},\mathtt{anc}\}$ denoting the BE ancilla,
the block encoding $\mathcal{O}_{\det}$ satisfies 
\begin{equation}
\left(\langle 0|_{\mathtt{sa}}\otimes \mathbb{1}_{\Phi}\right)
\mathcal{O}_{\det}
\left(|0\rangle_{\mathtt{sa}}\otimes \mathbb{1}_{\Phi}\right)
=
\frac{D_{\det}}{\alpha_{\det}},
\label{eq::det_oracle}
\end{equation}
where two properties of the normalization factor, $\alpha_{\det}$, in Eq.~\eqref{eq::det_oracle} are worth emphasizing. 
First, $\alpha_{\det}$ is independent of the auxiliary-field configuration by construction, so post-selection rescales all weights uniformly and never biases the prepared distribution. 
Second, its magnitude, which sets the entire post-selection cost, is determined by the underlying model and its parameters through the shape of the DQMC distribution; this dependence is analyzed in Sec.~\ref{sec:loading}. 
The post-selection overhead arises only in the probability-loading step.

We describe the construction of the determinant oracle in broad brushstrokes here, with the details deferred to Appendix~\ref{app:state_prep_oracle}. The construction uses block encoding (BE), linear combination of unitaries (LCU) and quantum singular value transformations (QSVT) as tools, and proceeds in four stages, summarized in Table~\ref{tab:state_prep_interface}: (i)~a block encoding of the action matrix $M_{\Phi}$ itself, in which the auxiliary-field register enters purely as control wires; this is where the free-fermion structure at fixed $\Phi$ is exploited, and why no enumeration of configurations ever occurs; 
(ii)~a QSVT polynomial that converts it into a block encoding of the matrix logarithm; 
(iii)~a coherent trace, evaluated as an amplitude on a maximally entangled system-copy pair, which compresses the matrix logarithm into the scalar log-weight per space-time site; 
and (iv)~a second QSVT polynomial that exponentiates this scalar.  The associated circuit depth and ancilla cost are also summarized in Table~\ref{tab:state_prep_interface}. Combining these four steps realizes the determinant oracle $\mathcal{O}_{\det} = \mathcal{U}_{\exp} \circ L_{\Phi} \circ \mathcal{U}_{\log} \circ \mathcal{U}_M$.

\begin{table*}[t]
\squeezetable
\caption{\label{tab:state_prep_interface} {\bf State-preparation pipeline}. The cost column reports stage-local query counts in terms of calls to the oracle in the preceding row. The resulting circuit depth for a single-call of the determinant-oracle is $\widetilde O(\kappa_{\log}N_{\tn{st}}^{3/2})$. The normalizations $\alpha_M$, $\alpha_{\log}$, and $\alpha_{\det}$ are derived quantities (See Appendix~\ref{app:state_prep_oracle}) rather than independently tuned hyperparameters. }
\begin{ruledtabular}
\begin{tabular}{lcccc}
\textbf{Opertation and Name} & \textbf{Outcome} &\textbf{Ancilla} & \textbf{Hyperparameter} & \textbf{Cost} \\
\hline
$\mathcal{U}_{M}$ : BE of $M_{\Phi}$
& BE of $M_{\Phi}/\alpha_M$
& $\mathtt{BE}$, $\mathtt{LCU}$ $\sim O(1)$ 
& None
& $O(N_{\mathrm{st}})$ in circuit-depth \\
$\mathcal{U}_{\log}$ : Log-QSVT
& BE of $\log\sqrt{M_{\Phi}^{\dagger}M_{\Phi}}/\alpha_{\log}$
& $\mathtt{anc_{QSVT}}$ $\sim O(1)$
& $\kappa_{\log}$, $\epsilon_{\log}$
& $O\!\left(\kappa_{\log}\log\epsilon_{\log}^{-1}\right)$ calls to $\mathcal{U}_M$ \\
$L_{\Phi}$: Coherent Trace
& Scalar block $\frac{1}{N_{\mathrm{st}}}
\log\left|\det \frac{M_{\Phi}}{\alpha_M}\right|$
& $\mathtt{sys'}$ $\sim O(\log N_{\mathrm{st}})$
& None
& $O(1)$ calls to $\mathcal{U}_{\log}$ \\
$\mathcal{U}_{\exp}$ : Exp-QSVT
& $|\det M_{\Phi}|/\alpha_{\det}$
& $\mathtt{anc_{QSVT'}}$ $\sim O(1)$
& $\alpha_{\mathrm{rescale}}$, $\epsilon_{\exp}$
& $O\!\left(\sqrt{N_{\tn{st}}\log\epsilon_{\exp}^{-1}}\right)$ calls to $L_{\Phi}$
\end{tabular}
\end{ruledtabular}
\end{table*}

The oracle synthesizes the determinant amplitudes coherently rather than loading classically precomputed probability data. The auxiliary-field register is initialized in the uniform superposition. Let $\mathtt{anc}$ denote the collection of all the ancillas in Table.~\ref{tab:state_prep_interface}. The BE ancilla is initialized in $\ket{0}_{\mathtt{sa}}$. The full qubit register is therefore initialized in the state
\begin{equation}
\ket{\mathrm{init}}=|0\rangle_{\mathtt{sa}}
|+\rangle_{\mathtt{aux}}\equiv \frac{1}{\sqrt{2^{N_{\tn{st}}}}}\sum_{\{\Phi\}}|0\rangle_{\mathtt{sa}}|\Phi\rangle_{\mathtt{aux}}.
\end{equation}
Applying $\mathcal{O}_{\det}$ gives
\begin{equation}
\mathcal{O}_{\det}\ket{\mathrm{init}}
=
A_M
|0\rangle_{\mathtt{sa}}
\sum_{\{\Phi\}}|\det M_{\Phi}|\,|\Phi\rangle_{\mathtt{aux}}
+|\perp\rangle,
\label{eq::det_oracle_action}
\end{equation}
where $|\perp\rangle$ has no support on the all-zero work-register subspace.  
Postselecting all work registers on $|0\rangle$ and normalizing therefore prepares Eq.~\eqref{eq:qalg_target}. The factor $A_M =\alpha_{\det}^{-1}$ controls the post-selection overhead. The selected amplitude carries the first power $|\det M_{\Phi}|\propto\sqrt{p_{\Phi}}$, while the Born rule supplies the second power required by the physical two-flavor distribution.

\subsection{Observable Measurement via Coherent Averaging}\label{sec:qalg_obs}
 
We next describe how observables are measured once the auxiliary-field superposition has been prepared. For each fixed HS configuration, the interacting problem reduces to free fermions, so the observables relevant to DQMC can be expressed in terms of single-particle Green's functions. In particular, after Wick decomposition, a broad class of equal-time and time-displaced correlation functions reduce to bilinear expressions of the convolution form
\begin{equation}
O_{\Phi}(q)
=
\frac{1}{N}
\sum_{rr'}
e^{iq\cdot(r-r')}
\left[A_{\Phi}\right]_{rr'}
\left[B_{\Phi}\right]_{rr'},
\label{eq:qalg_Ophi}
\end{equation}
where $q\equiv(\Omega_n,\vec{q})$ and $r\equiv (\tau,\vec{r})$ are the 3-momentum and 3-coordinate, $N=N_{\tn{st}}$ is the space-time volume of the convolution, and $A_{\Phi},B_{\Phi}$ are the two Wick kernels (built from configuration-resolved Green's functions) of a fermion-bilinear correlator conditioned on $\Phi$, respectively. For a correlator of a fermion bilinear with its Hermitian conjugate, $O_{\Phi}(q)$ reduces to the imaginary-time susceptibility $\chi_{O}(\vec q,i\Omega_n)$ of Eq.~\eqref{eq:def_corr}.

The kernels $A_{\Phi}$ and $B_{\Phi}$ are obtained from the same action-matrix oracle $\mathcal{U}_{M}$ in Table~\ref{tab:state_prep_interface} used in state preparation. The Green's function $G_{\Phi}\equiv M_{\Phi}^{-1}$ is realized by applying a QSVT polynomial approximating $1/x$ on the singular-value interval of $M_{\Phi}/\alpha_M$~\cite{GilyenSuLowWiebe2019QSVT,HHL2009}; approximating $1/x$ to precision $\epsilon$ on $[\kappa_G^{-1},1]$ requires degree $O(\kappa_G\log\epsilon^{-1})$, where $\kappa_G$ is the condition number of $M_{\Phi}$. The resulting circuits block-encode the inverse kernels (the imaginary-time-ordered Green's functions) needed for the observable. We denote these block encodings by $\mathcal{U}_{A}(\Phi)$ and $\mathcal{U}_{B}(\Phi)$, with normalizations $\alpha_A$ and $\alpha_B$, defined as
\begin{equation}
\begin{aligned}
\left(\langle 0|_{\mathrm{BE_0}}\otimes \mathbb{1}\right)
\mathcal{U}_{A}(\Phi)
\left(|0\rangle_{\mathrm{BE_0}}\otimes \mathbb{1}\right)
&=
\frac{A_{\Phi}}{\alpha_A},
\\
\left(\langle 0|_{\mathrm{BE_1}}\otimes \mathbb{1}\right)
\mathcal{U}_{B}(\Phi)
\left(|0\rangle_{\mathrm{BE_1}}\otimes \mathbb{1}\right)
&=
\frac{B_{\Phi}}{\alpha_B}.
\end{aligned}
\label{eq:qalg_BE_AB}
\end{equation}
 
To evaluate the correlation function at a specific 3-momentum $q$, we use a convolutional quantum Fourier transform (cQFT) acting on a system-copy pair, $\mathtt{sys}$ and $\mathtt{sys}'$. A gate-level realization is shown in Fig.~\ref{Fig::fig_fullq_gates_main}(b), denoted as cQFT-PREP. Starting from a 3-momentum label $q$ on register $\mathtt{sys}$ and the zero state on $\mathtt{sys'}$, the cQFT prepares the coherent pair state
\begin{equation}
|q;0\rangle
\longmapsto
\frac{1}{\sqrt{N}}
\sum_m
e^{iq\cdot r_m}
|m,m\rangle.
\label{eq:qalg_cqft}
\end{equation}
Applying the two block encodings of Eq.~\eqref{eq:qalg_BE_AB} to the system-copy pair and then applying the inverse cQFT gives the composite unitary
\begin{equation}
\mathcal{W}_{AB}(\Phi)
\equiv
\mathcal{U}_{\mathrm{cQFT}}^{\dagger}
\left(\mathcal{U}_{A}(\Phi)\otimes \mathcal{U}_{B}(\Phi)\right)
\mathcal{U}_{\mathrm{cQFT}} ,
\label{eq:qalg_WAB}
\end{equation}
whose selected matrix element yields the configuration-resolved observable,
\begin{equation}
\langle 0,0,q,0|
\mathcal{W}_{AB}(\Phi)
|0,0,q,0\rangle
=
\frac{O_{\Phi}(q)}{\alpha_A\alpha_B},
\label{eq:qalg_cqft_amp}
\end{equation}
where the two leading zero registers denote the block-encoding ancillas and the last two denote $\mathtt{sys}$ and $\mathtt{sys'}$. For a general complex $O_{\Phi}$, its real and imaginary parts can be extracted separately by the two variants of the Hadamard test below.
 
At this point there are two conceptually distinct strategies. If $O_{\Phi}(q)$ is per-configuration positive-semidefinite, one may first measure the auxiliary register, collapsing the system onto a definite HS configuration $\Phi$, evaluate $O_{\Phi}(q)$ via Eq.~\eqref{eq:qalg_cqft_amp}, and average over configurations classically; as discussed in Sec.~\ref{sec:compat} below. The principal advantage of the full quantum algorithm, however, is that the average can be performed \emph{coherently} via a Hadamard test.
 
A gate-level construction of the Hadamard test is shown in Fig.~\ref{Fig::fig_fullq_gates_main}(c). With the $\mathrm{DQ}^{2}\mathrm{MC}$ state of Eq.~\eqref{eq:qalg_target} prepared on $\mathtt{aux}$ as described in Sec.~\ref{sec:qalg_state_prep}, we introduce a Hadamard-test ancilla $\mathtt{anc}_H$ and use it to control the gate set that block-encodes the observable oracles of Eq.~\eqref{eq:qalg_BE_AB}. The interference term then carries the ensemble average, and the probability of the ancilla reading $|0\rangle$ is
\begin{equation}
p(0_{\mathrm{H}})
=
\frac{1}{2}
\;+\;
\frac{1}{2\alpha_A\alpha_B}
\operatorname{Re}\!\left[
\overline{O}(q)
\right],
\label{eq:qalg_hadamard}
\end{equation}
where
\begin{equation}
\overline{O}(q)
\equiv
\sum_{\{\Phi\}}p_{\Phi} O_{\Phi}(q),
\label{eq:qalg_Oavg}
\end{equation}
and the imaginary part follows from the standard phase-shifted variant of the same test. Equation~\eqref{eq:qalg_hadamard} makes the coherent averaging explicit: once $|\mathrm{DQ}^{2}\mathrm{MC}\rangle$ is prepared, the cost is set by the precision of a single interference experiment. The number of experiments required is subject to the standard central limit theorem. Resolving $\overline{O}(q)$ to additive error $\epsilon$ therefore costs $O((\alpha_A\alpha_B/\epsilon)^{2})$ Hadamard-test shots. This factor enters the end-to-end runtime estimate of Sec.~\ref{sec:qalg_summary}, where each Hadamard test requires a freshly prepared state.
 
We close with the finite-size scaling, which makes the distinction between ordered and disordered channels transparent. Suppose $O$ develops long-range order at $\q=0$, so that the \emph{$\Phi$-averaged} correlator satisfies $\langle O_{\r}O_{\r'}\rangle=\lambda_{O}$ independent of $|\r-\r'|$, with $\lambda_O$ the order parameter surviving the auxiliary-field average. The averaged kernel then carries a single macroscopic eigenvalue $\sim N_{s}\lambda_{O}$ ($N_{s}=L^{2}$ the spatial volume); while the block-encoding normalization $\alpha_A\alpha_B$ scales as $N_{s}$ and absorbs this factor, since the QSVT-inverse normalization inherits the spatial dimension of $M_{\Phi}$ through its condition number. Substituting into Eq.~\eqref{eq:qalg_hadamard} gives $p(0_{\mathrm{H}})-1/2\to \lambda_{O}/2$, independent of $N_{s}$: the ordered channel is precision-efficient, requiring $O(1/\epsilon^{2})$ shots with additive error $\epsilon$ in observable estimation. The analysis with long-range order at $\q=0$ can carry through analogously to a finite-$\q$ order. In contrast, an unordered channel has only $O(1)$ kernel eigenvalues, $p(0_{\mathrm{H}})-1/2$ decays as $1/N_{s}$, and resolving it costs $O(N_s^{2}/\epsilon^{2})$ shots. The deviation of $p(0_{\mathrm{H}})$ from $1/2$, and its scaling with $N_s$, thus serves as a direct diagnostic of long-range order at $\q$ This shows the full quantum protocol can analyze phases of matters via finite-size scaling of correlation functions, as is done within DQMC.

\subsection{Postselection and Hyperparameter Scaling}\label{sec:loading}

\begin{figure}[htb!]
\centering
\includegraphics[width=8cm]{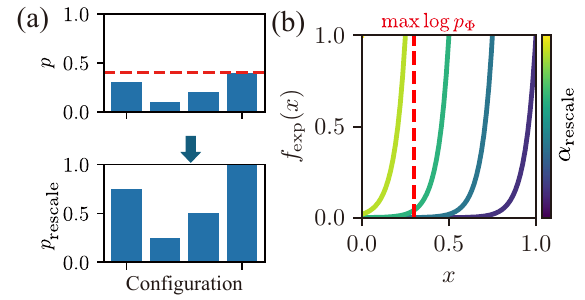}
\caption{\label{Fig::precon} {\bf Schematic of Preconditioning}  (a) Schematic of the rescaling: the bare weight distribution $p_{\Phi}$ (top) is rescaled by $\alpha_{\mathrm{rescale}}$ so that the largest weight saturates the block-encoding bound (bottom). (b) Schematics of clipped target function in Exp-QSVT, tuned by $\alpha_{\tn{rescale}}$. Since the signal domain for QSVT is the unit interval, regions where the function value $>1$ are clipped to $1$. To maintain intact probability weight, $\alpha_{\tn{rescale}}$ should be chosen such that the clipping is inactive for the $\max\log p_{\Phi}$, denoted by red dashed line.}
\end{figure}

We refer to the action of the determinant oracle on the initial state as probability-loading.  
The probability-loading cost is controlled by two hyperparameters: a configuration-independent weight rescaling factor $\alpha_{\tn{rescale}}$ controlling the success probability, and $\kappa_{\log}$ introduced in Log-QSVT controlling the circuit depth. 

First, the success probability of post-selection $p_{\mathrm{succ}}$ is governed by the amplitude $A_M$ in Eq.~\eqref{eq::det_oracle_action}, given by
\be
\label{eq:def_Am}
A_M = \frac{e^{N_{\tn{st}}x_0}}{\alpha_M^{N_{\tn{st}}} \sqrt{2^{N_{\tn{st}}}}}.
\ee
Here, $x_0$ controls the configuration-independent weight rescaling, parameterized by $\alpha_{\mathrm{rescale}}$ as
\begin{equation}
x_0\equiv \log(\alpha_{\mathrm{rescale}})+\log(\alpha_{M}).
\label{eq:qalg_x0}
\end{equation}
The $\log\alpha_M$ term cancels the action-matrix block-encoding normalization, whereas $\alpha_{\mathrm{rescale}}$ rescales the encoded weight globally for all the configurations. The optimal rescaling is reached when the largest encoded weight saturates unity, as schmatically shown in Fig.~\ref{Fig::precon}(a). 
At this optimum, the success probability is
\begin{equation}
p_{\mathrm{succ}}
=
\frac{1}{2^{N_{\tn{st}}}\,p_{\max}}
\;\ge\;
\frac{p_{\min}}{p_{\max}}
\equiv
\kappa^{-1},
\label{eq:qalg_psucc}
\end{equation}
where $\kappa\equiv p_{\max}/p_{\min}$ is the condition number of the DQMC distribution in Eq.~\eqref{eq:qalg_pphi}. Thus, $p_{\mathrm{succ}}$ is polynomial in $1/N_{\tn{st}}$ precisely when $p_{\max}\lesssim \tn{poly}(N_{\tn{st}})/2^{N_{\tn{st}}}$; it can be exponentially small for a sharply peaked distribution.

This degree of freedom to rescale comes from the implementation of Exp-QSVT; see Table ~\ref{tab:state_prep_interface}. Since the signal domain of QSVT is the unit interval, the target exponential function $f_{\exp}(x)$ should be clipped, where values larger than $1$ are set to $1$, i.e. $f_{\exp}(x)=\min\!\left\{1,e^{N_{\mathrm{st}}(x+x_0)}\right\}$. The variable $x$ here denotes the log-weight of the DQMC distribution. To faithfully implementing the intact probability weight, $\alpha_{\tn{rescale}}$ should be chosen such that for the largest log-weight the clipping window is still inactive [Fig.~\ref{Fig::precon}(b)]. 
Within the faithful, unclipped window of the QSVT implementation, $p_{\mathrm{succ}}\propto\alpha_{\mathrm{rescale}}^{2N_{\tn{st}}}$.

The second hyperparameter, $\kappa_{\tn{log}}$, sets the singular-value interval retained by log-QSVT. Only configurations whose probability relative to the maximum is $\geq\kappa_{\log}^{-2}$ are unaffected. Thus, keeping the entire probability range intact requires $\kappa_{\tn{log}}\gtrsim\sqrt{\kappa}$. A sharply peaked distribution can require an exponentially large $\kappa_{\log}$. Since the log-QSVT degree is $O\!\left(\kappa_{\log}\log\epsilon_{\log}^{-1}\right)$, $\kappa_{\log}$ can produce an exponential circuit depth at worst; a smaller cutoff may be permissible when the total discarded probability lies below the target infidelity.

A benchmark as a proof-of-demonstration for the Hubbard model at small system-size and high temperature (on a $2\times2\times2$ space-time lattice) can be found in Appendix~\ref{app:hubbard_benchmarks}, together with a detailed analysis of preconditioning hyperparameters.

\subsection{Summary and Resource Scaling}\label{sec:qalg_summary}
The quantum advantage of the fully quantum protocol is twofold.
First, both the state preparation and the observable measurement are
performed coherently, so the issue of autocorrelation time simply does not arise; no sampling ever occurs. 
Second, the circuit-depth of a single oracle call is therefore $\widetilde{O}(\kappa_{\tn{log}}N_{\tn{st}}^{3/2})$, where $\widetilde{O}$ suppresses logarithmic factors. 
In contrast, conventional DQMC costs $O(L_{\tn{T}}N_s^{3})$ per sweep. With $N_{\tn{st}}=L_{\tn{T}}N_s$, the quantum oracle depth $\widetilde O(N_{\tn{st}}^{3/2})$ scales more steeply with the temporal extent but  mildly with spatial volume, before including the contributions from $p_{\mathrm{succ}}^{-1}$ and $\kappa_{\log}$. The extra factor $N_{\tn{st}}^{1/2}$ beyond the linear action-matrix-oracle depth arises specifically from the degree of the exp-QSVT approximation (see Appendix~\ref{app:exp_qsvt_degree}).
 
Since each Hadamard-test shot consumes a freshly prepared
$\ket{\mathrm{DQ}^{2}\mathrm{MC}}$ state, it is worth assembling the
end-to-end cost of estimating one observable explicitly. Preparing the
state requires on average $p_{\mathrm{succ}}^{-1}$ oracle calls
(improvable to $p_{\mathrm{succ}}^{-1/2}$ by amplitude
amplification~\cite{Brassard2002AA}), and resolving the interference
signal of Eq.~\eqref{eq:qalg_hadamard} to additive error $\epsilon$
requires $N_{\mathrm{shot}}=O\!\left((\alpha_A\alpha_B/\epsilon)^{2}\right)$
repetitions, where $\alpha_A,~\alpha_B$ were introduced in Eq.~\eqref{eq:qalg_BE_AB}. The total runtime is therefore
\begin{equation}
T_{\tn{full}}
\sim
\widetilde{O}\!\left(
\kappa_{\tn{log}}
\frac{N_{\tn{st}}^{3/2}}{p_{\mathrm{succ}}}
\left(\frac{\alpha_A\alpha_B}{\epsilon}\right)^{2}
\right),
\label{eq:qalg_total_cost}
\end{equation}
where $\widetilde{O}$ suppresses logarithmic factors.
The observable-dependent factor also matters: by the finite-size analysis
of Sec.~\ref{sec:qalg_obs}, the block-encoding normalization
$\alpha_A\alpha_B\propto N_s$ is compensated by a macroscopic kernel
eigenvalue for long-range-ordered channels for which the protocol
is precision-efficient; whereas unordered channels carry a signal
suppressed as $1/N_s$ and correspondingly larger shot counts.
 
The remaining and dominant, uncertainty in
Eq.~\eqref{eq:qalg_total_cost} is $p_{\mathrm{succ}}$ and $\kappa_{\tn{log}}$, which is set by
the condition number $\kappa$ of the underlying DQMC distribution
[Eq.~\eqref{eq:qalg_psucc}]. This bottleneck is fundamental, in a
sense that goes beyond the probability loading and synthesizing task itself. From the
viewpoint of QITE~\cite{Motta2020QITE,McArdle2019VITE}, the log-QSVT procedure
effectively prepares a block encoding of a diagonal ``Hamiltonian''
whose Hilbert-space dimension is $N_{\tn{eff}}\equiv 2^{N_{\tn{st}}}$,
and the exp-QSVT then targets a purified thermal state of this
Hamiltonian at effective inverse temperature
$\beta_{\tn{eff}}\equiv N_{\tn{st}}$. To the best of our knowledge, no
quantum algorithm prepares such state,or more generally,
the (mixed) thermal state of a generic Hamiltonian
deterministically at cost
$O(\log N_{\tn{eff}})$~\cite{PoulinWocjan2009Gibbs,Temme2011Metropolis,ChowdhurySomma2017Gibbs}.
Amplitude amplification improves $p_{\mathrm{succ}}$ only
quadratically, so when $p_{\mathrm{succ}}$ is exponentially small in
$N_{\tn{st}}$ the overall preparation cost remains exponential.

\begin{table}[t]
\caption{{\bf Qubit counts for the fully quantum state preparation protocol.} Here, $N_{\tn{st}}$ denotes the space-time volume, $n_{\tn{orb}}$ denotes the number of orbitals per site per flavor, and $n_{\tn{vert}}$ denotes the number of interaction vertices per site per flavor. Both $n_{\tn{orb}}$ and $n_{\tn{vert}}$ are $O(1)$ numbers.}
\label{tab:fullq_qubit_count}
\begin{ruledtabular}
\begin{tabular}{lc}
Qubit Register & Number Count \\
\hline
$\mathtt{aux}$: auxiliary field qubits & $n_{\tn{vert}} N_{\tn{st}}$ \\
$\mathtt{sys}$: system qubits for the action-matrix & $\log(n_{\tn{orb}}N_{\tn{st}})$ \\
$\mathtt{sys}'$: system copy for trace estimation & $\log(n_{\tn{orb}}N_{\tn{st}})$ \\
$\mathtt{BE}$: BE ancilla for the action-matrix  & $1$ \\
$\mathtt{LCU}$: LCU ancilla for the action-matrix  & $1$ \\
$\mathtt{anc_{QSVT}}$: QSVT ancilla for logarithm  & $1$ \\
$\mathtt{anc_{QSVT}'}$: QSVT ancilla for exponentiation  & $1$ \\
\hline
Total & \hspace*{-4.75em}$n_{\tn{vert}}N_{\tn{st}} +2\log(n_{\tn{orb}}N_{\tn{st}}) + 4$
\end{tabular}
\end{ruledtabular}
\end{table}
 
Finally, the qubit count is favorable, benefiting from the fact that
at fixed configuration the system is effectively non-interacting,
the action matrix has dimension $N_{\tn{st}}$ rather than
$2^{N_s}$. Only $\log (N_{\tn{st}}) + O(1)$ system qubits are needed
to block-encode the action matrix; coherent trace estimation requires
one ancillary copy of the system register, and no other ancilla count
scales with $N_{\tn{st}}$.
The full state-preparation stage uses one auxiliary qubit per binary HS variable and hence $O(N_{\tn{st}})$ auxiliary qubits, resulting in a total qubit count of $O(N_{\tn{st}})$. 
Specifically, for a generic model satisfying the assumptions in Sec.~\ref{sec:qalg} with $n_{\tn{vert}}$ interaction vertices and $n_{\tn{orb}}$ orbitals per site per flavor, a detailed summary of qubit counts for the fully quantum state preparation protocol is in Table~\ref{tab:fullq_qubit_count}.
As we show in Sec.~\ref{sec:compat}, the hybrid
quantum-classical protocol reduces the auxiliary-qubit count to a
flexibly chosen constant between $1$ and $N_{\tn{st}}$ depending on the updating scheme, at the price of a classical sampling loop over
auxiliary-field configurations. 
In this work we specialize to a $O(N_{s})$ active block size, corresponding to a spatial global update.
We place these features in the
context of prior quantum approaches to Monte Carlo in
Sec.~\ref{sec:related} and Table~\ref{tab:comparison}.

\section{Hybrid Quantum-Classical Protocol}\label{sec:compat}
 
We now demonstrate the compatibility of the $\mathrm{DQ}^{2}\mathrm{MC}$ algorithm with a hybrid quantum-classical setup. The key is to downgrade part of the auxiliary-field register from quantum to classical, and to generate classical samples of the auxiliary field in a Monte-Carlo fashion by sweeping the quantum active block across the space-time lattice. In sharp contrast with classical DQMC, this construction naturally allows global updates with no additional oracle cost per update, as we explain below.

\begin{figure*}[htb!]
\includegraphics[width=175mm,scale=0.5]{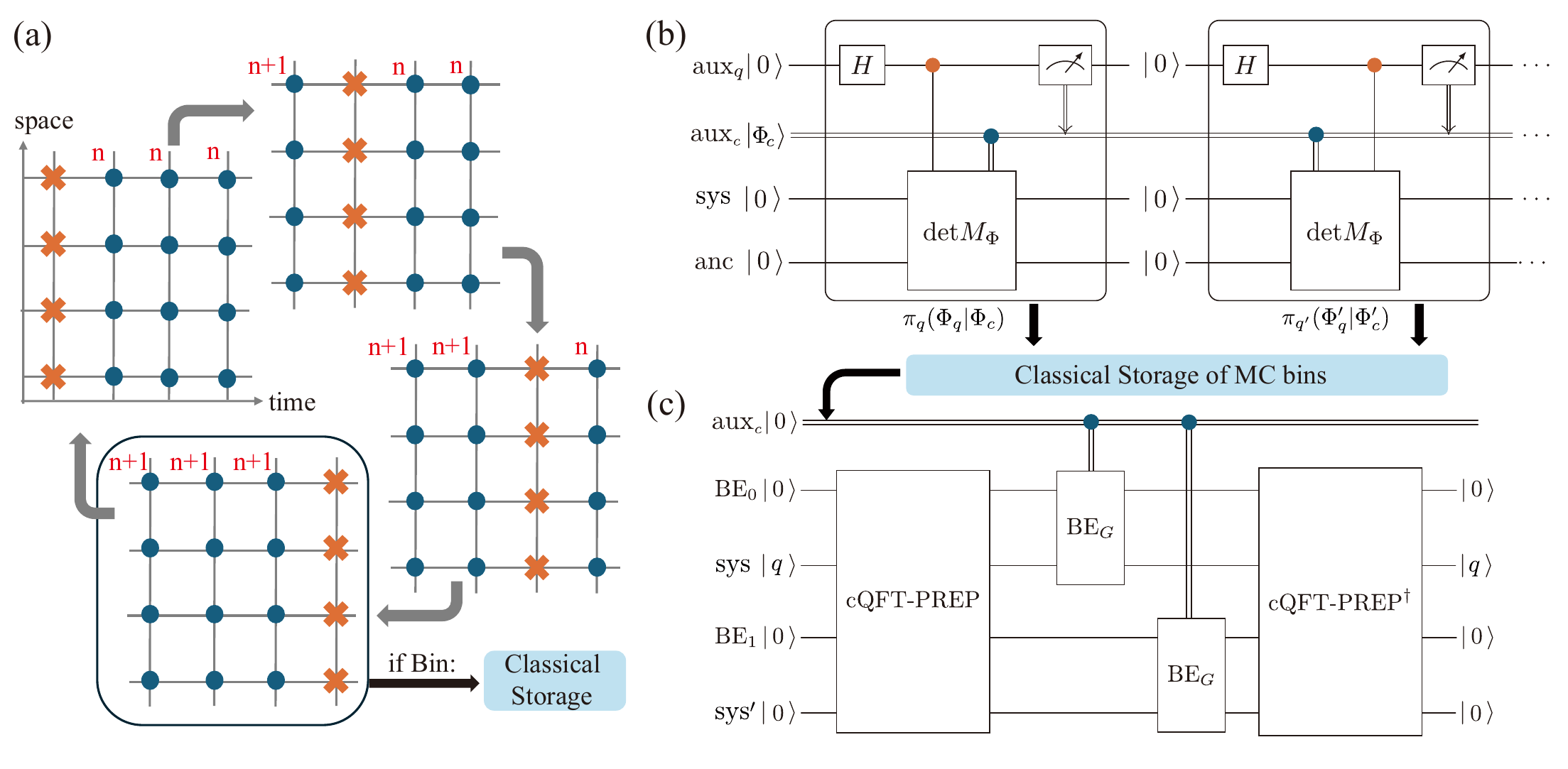}
\caption{\label{Fig::fig3} {\bf Hybrid quantum-classical protocol.} (a) Block sweep on the space-time auxiliary field: at Monte-Carlo step $n$ the active block (orange crosses) is kept quantum, while the remaining fields (blue dots) are fixed as classical controls on the action-matrix oracle $\mathcal{U}_{M}$; the active block is updated by measurement and the resulting bits replace the corresponding entries of the classical configuration. When the step is retained as a bin, the full classical configuration $\Phi_{c}$ is passed to the observable circuit. (b) Hybrid update circuit: $\mathcal{O}_{\det}$ (boxed gate $\det M_{\Phi}$) is applied to the Hadamard-prepared active register $\mathtt{aux}_{q}$ while the complementary register $\mathtt{aux}_{c}$ carries the classical $\Phi_{c}$; measurement of $\mathtt{aux}_{q}$ samples $\pi_{q}(\Phi_{q}|\Phi_{c})$. Successive updates alternate the active block by rewiring. (c) Measurement of observables which are per-configuration sign definite [Eq.~\eqref{eq:hyb_obs_def}]: $\mathtt{aux}_{c}$ carries $\Phi_{c}$ and controls the two Green's-function block encodings $\mathrm{BE}_{G}$, acting on $\mathtt{sys},\mathtt{sys}'$ prepared in the coherent momentum-pair state by the cQFT-PREP block.}
\end{figure*}
 
\subsection{Cluster-Gibbs Sampler}\label{subsec:gibbs}
 
The determinant oracle $\mathcal{O}_{\det}$ of Sec.~\ref{sec:qalg} is a block-encoded diagonal filter on the auxiliary-field register. In the fully coherent preparation it acts on the uniform state $|+\rangle_{\mathtt{aux}}$.
Partitioning the auxiliary fields on the space-time lattice into a quantum active block $\Phi_{q}$ and a classical complement $\Phi_{c}$, i.e. $\Phi=(\Phi_{q},\Phi_{c})$, allows the same state-preparation primitive to be run in a traditional Monte-Carlo fashion.
The number of auxiliary qubits $|\Phi_{q}|$ can be chosen flexibly depending on the updating scheme: $|\Phi_{q}|=1$ corresponds to single-spin-flip updates, and $1<|\Phi_{q}|<N_{\tn{st}}$ corresponds to a global (cluster) update.
To perform the sampling, the active block $\Phi_{q}$ is swept across the space-time lattice according to a schedule that collectively covers all auxiliary variables.
An example corresponding to a spatial global update is shown in Fig.~\ref{Fig::fig3}(a), where $\Phi_{q}$ is a single time-slice and the schedule is a forward-and-backward sweep across the time direction.
In one update, only the registers of the active block $\Phi_{q}$ are updated, while the complement $\Phi_{c}$ is cached classically and treated as classical controls on the HS-dependent one-body vertices $U_{\gamma,i}$ of $\mathcal{U}_{M}$ constructed in Appendix~\ref{app:state_prep_oracle}. Moving from one update to the next only requires rewiring the circuit layout of $\mathcal{U}_{M}$ according to the new assignment of $\Phi_{q}$ [see Fig.~\ref{Fig::fig3}(b)], while the rest of the circuit structure is unchanged.
 
To start the algorithm, the classical controls $\Phi_{c}$ are picked randomly, and the active block is initialized in
\begin{equation}
|+\rangle_{q}
=
\frac{1}{\sqrt{2^{|{\Phi_{q}|}}}}
\sum_{\{\Phi_{q}\}}|\Phi_{q}\rangle,
\end{equation}
where $|\Phi_{q}|$ is the number of active HS variables. As in Eq.~\eqref{eq::det_oracle_action}, postselecting the work registers on $|0\rangle_{\mathtt{sa}}$ loads the amplitude corresponding to the conditional distribution of the active block at fixed $\Phi_{c}$,
\begin{subequations}
\beq
|+\rangle_{q}
&\rightarrow&
\frac{1}{\sqrt{\mathcal{N}(\Phi_{c})}}
\sum_{\{\Phi_{q}\}}
|\det M(\Phi_{q},\Phi_{c})|\,|\Phi_{q}\rangle, ~~~~~\\
\mathcal{N}(\Phi_{c})
&=&
\sum_{\{\Phi_{q}\}}
|\det M(\Phi_{q},\Phi_{c})|^{2},
\label{eq:compat_cond_state}
\eeq
\end{subequations}
where $\mathcal{N}(\Phi_{c})$ is the conditional normalization, proportional to the marginal weight $p(\Phi_{c})$.
A computational-basis measurement of the active register $\Phi_{q}$ then returns samples from the conditional DQMC distribution,
\begin{equation}
\pi_{q}(\Phi_{q}|\Phi_{c})
=
\frac{p_{\Phi}}{\sum_{\{\Phi'_{q}\}}p_{(\Phi'_{q},\Phi_{c})}},
\label{eq:compat_cond_prob}
\end{equation}
with $p_{\Phi}$ the physical two-flavor DQMC probability of Eq.~\eqref{eq:qalg_pphi}, built from the squared magnitude of the single-flavor determinant. Preparation and measurement of the active block therefore realize a heat-bath update of $\Phi_{q}$ at fixed $\Phi_{c}$ [Fig.~\ref{Fig::fig3}(b)].
To advance, the measurement outcome for $\Phi_{q}$ is cached classically and becomes part of the new $\Phi_{c}$ fed into $\mathcal{U}_{M}$; the quantum register $\mathtt{aux}_q$ is re-initialized in $|+\rangle$ for the next update but wired to control the $U_{\gamma,i}$ according to the schedule [Fig.~\ref{Fig::fig3}(a)].
Iterating over a schedule of active blocks $q_{1},q_{2},\ldots$ that collectively covers the space-time lattice generates a cluster-Gibbs sweep over $\Phi$. Once enough sweeps have been performed, the resulting $\Phi$ configurations in the classical cache are collected and stored for the observable-measurement stage.

The Markov kernel of one block update,
\begin{equation}
K_{q}(\Phi\rightarrow\Phi')
=
\pi_{q}(\Phi'_{q}|\Phi_{c})\,
\delta_{\Phi'_{c},\Phi_{c}},
\label{eq:compat_kernel}
\end{equation}
is the block-Gibbs (heat-bath) kernel on the active block $q$: the new block is drawn directly from the conditional target, so no Metropolis-Hastings acceptance step is required. Using $p_{\Phi}=p(\Phi_{c})\,\pi_{q}(\Phi_{q}|\Phi_{c})$,
\begin{equation}
p_{\Phi}\,K_{q}(\Phi\rightarrow \Phi')
=
p(\Phi_{c})\,\pi_{q}(\Phi_{q}|\Phi_{c})\,\pi_{q}(\Phi'_{q}|\Phi_{c})
\delta_{\Phi'_{c},\Phi_{c}},
\label{eq:compat_db}
\end{equation}
which is symmetric under $\Phi_{q}\leftrightarrow\Phi'_{q}$ on the support of the kernel; thus each fixed-block update satisfies detailed balance with respect to $p_{\Phi}$.
Composing updates along a schedule preserves $p_{\Phi}$. Because the heat-bath kernel assigns nonzero probability to every value of the active block whenever the corresponding weights are nonzero, a schedule that covers every HS variable yields an irreducible, aperiodic chain, so $p_{\Phi}$ is its unique stationary distribution.
 
Each cluster-Gibbs update inherits the post-selection analysis of Sec.~\ref{sec:loading}, applied to the conditional distribution. Defining the condition number of the conditional distribution on the active block,
\begin{equation}
\kappa_{q}(\Phi_{c})
\equiv
\frac{\max_{\Phi_{q}}\pi_{q}(\Phi_{q}|\Phi_{c})}
{\min_{\Phi_{q}}\pi_{q}(\Phi_{q}|\Phi_{c})},
\label{eq:compat_kappa_q}
\end{equation}
the same preconditioning that tightened Eq.~\eqref{eq:qalg_psucc} gives, at optimal rescaling, the conditional success probability
\begin{equation}
p_{\mathrm{succ}}(\Phi_{c})
=
\frac{1}{2^{|\Phi_{q}|}\,\max_{\Phi_{q}}\pi_{q}(\Phi_{q}|\Phi_{c})}
\;\ge\;
\kappa_{q}(\Phi_{c})^{-1}.
\label{eq:compat_psucc}
\end{equation}
In the worst case this decays exponentially with the active-block size $|\Phi_{q}|$. Smaller blocks therefore have higher success rates but require more local moves per sweep; larger blocks reduce the Markov-chain autocorrelation time but demand more coherent auxiliary qubits per update. The costs for the hybrid protocol are summarized in Table~\ref{tab:hyb_cost}. 

\begin{table}[t]
\caption{{\bf Hybrid resource scaling.} Here $u$ labels block updates, $T_{\mathrm{sweep}}$ is the depth cost per sweep, and $\tau_{\mathrm{auto}}$ the auto-correlation time is measured in the number of sweeps.}
\label{tab:hyb_cost}
\begin{ruledtabular}
\begin{tabular}{lc}
Quantity & Expected circuit depth \\
\hline
One oracle attempt & $\widetilde O(\kappa_{\log}N_{\tn{st}}^{3/2})$ \\
Accepted update $u$ & $\widetilde O(\kappa_{\log}N_{\tn{st}}^{3/2}/p_{\mathrm{succ},u})$ \\
One sweep & $\widetilde O\!\left(\kappa_{\log}N_{\tn{st}}^{3/2}\sum_{u}p_{\mathrm{succ},u}^{-1}\right)$ \\
Independent bin & $O\!\left(\tau_{\mathrm{auto}}T_{\mathrm{sweep}}\right)$ \\
\end{tabular}
\end{ruledtabular}
\end{table}

A structural feature of the hybrid scheme deserves emphasis: because only classical information passes between successive updates, a failed post-selection at one update can be retried \emph{in place}, without restarting the chain. The costs of successive updates therefore \emph{add} rather than multiply. The expected number of oracle calls per sweep is proportional to $\sum_{u}p_{\mathrm{succ},u}^{-1}$ where $u$ labels block updates; this is in sharp contrast to the full-quantum protocol, where a single global post-selection failure discards the entire preparation. This additivity is what makes the hybrid protocol robust to a per-update success probability that would be prohibitive if it had to be paid coherently and simultaneously across the whole lattice. It also enables parallelization: independent walkers can be run on separate quantum processing units with parallel tempering among them. Each temperature running its own oracle wiring, and exchanging only classical configurations, which further reduces the autocorrelation time and admits a distributed implementation with purely classical inter-node communication, increasing the practicality of the approach in the NISQ era. The natural hardware target is therefore an early fault-tolerant platform with fast mid-circuit measurement and reset, low-latency classical feed-forward, and multiple processors capable of running independent determinant-oracle circuits; coherent quantum links between nodes are not required. Specifically, for a generic system with space-time volume $N_{\tn{st}}$ and with $n_{\tn{orb}}$ orbitals per site per flavor, the single spin-flip update only requires $2\log(n_{\tn{orb}}N_{\tn{st}}) + 5$ qubits, with a detailed summary of the qubit counts in Table~\ref{tab:hyb_qubit_count}. Let us use a DQMC study done by the authors \cite{wang2024prb, wang2025skyrmion} as an explicit example for comparison, where the finite temperature run with the largest space-time volume has $L=12$ in two-dimensions with $4$-orbitals per site per flavor and $N_{T}=300$ Trotter steps. Performing a DQ$^2$MC simulation on a quantum-classical hybrid hardware using single spin-flip update requires $41$ qubits.

\begin{table}[t]
\caption{{\bf Qubit counts for the hybrid state preparation with single spin-flip update.} Here, $N_{\tn{st}}$ denotes the space-time volume, and $n_{\tn{orb}}$ denotes the number of orbitals per site. $n_{\tn{orb}}$ is a $O(1)$ number.}
\label{tab:hyb_qubit_count}
\begin{ruledtabular}
\begin{tabular}{lc}
Qubit Register & Number Count \\
\hline
$\mathtt{aux}$: auxiliary field qubits & $1$ \\
$\mathtt{sys}$: system qubits for the action-matrix & $\log(n_{\tn{orb}}N_{\tn{st}})$ \\
$\mathtt{sys}'$: system copy for trace estimation & $\log(n_{\tn{orb}}N_{\tn{st}})$ \\
$\mathtt{BE}$: BE ancilla for the action-matrix  & $1$ \\
$\mathtt{LCU}$: LCU ancilla for the action-matrix  & $1$ \\
$\mathtt{anc_{QSVT}}$: QSVT ancilla for logarithm  & $1$ \\
$\mathtt{anc_{QSVT}'}$: QSVT ancilla for exponentiation  & $1$ \\
\hline
Total & $2\log(n_{\tn{orb}}N_{\tn{st}}) + 5$
\end{tabular}
\end{ruledtabular}
\end{table}
 
\subsection{Observable Measurement}\label{subsec:hybrid_obs}

We estimate observables from the stored configurations using the same composite unitary $\mathcal{W}_{AB}(\Phi)$ defined in Eq.~\eqref{eq:qalg_WAB}; the only difference from the full-quantum protocol is that the HS configuration is supplied as classical controls rather than coherently averaged over an auxiliary register. For a stored configuration $\Phi^{(c)}$, the ordinary and phase-shifted Hadamard tests of Sec.~\ref{sec:qalg_obs} give unbiased estimates $\widehat{O}_{\Phi^{(c)}}(q)$ of the real and imaginary parts of $O_{\Phi^{(c)}}(q)$. The hybrid estimator is
\begin{equation}
\widehat{\overline{O}}_{\tn{hyb}}(q)
=
\frac{1}{N_{\tn{bin}}}
\sum_{c=1}^{N_{\tn{bin}}}
\widehat{O}_{\Phi^{(c)}}(q),
\label{eq:compat_hyb_obs}
\end{equation}
where $\widehat{~}$ denotes Hadamard test estimator per configuration and $\overline{~}$ denotes the classical averaging.
Resolving a per-configuration amplitude to additive error $\epsilon$ requires $O((\alpha_A\alpha_B/\epsilon)^2)$ Hadamard-test shots, while the classical average has the usual variance governed by $N_{\tn{bin}}$ and the issue of the autocorrelation time. A benchmark for the Hubbard model can be found in Appendix~\ref{app:hubbard_benchmarks}.

Let us introduce a different approach to estimate the observable under a positivity condition, stated below.
When $O_{\Phi}(q)$ is independently known to be real with a common known sign across all configurations, the protocol in Fig.~\ref{Fig::fig3}(c) provides an alternative with fewer ancillas. After applying the inverse of cQFT-PREP gate and measuring all registers, the conditional success probability is
\begin{equation}
p(00_{\tn{BE}}\,|\,q_{\tn{sys}},0_{\tn{sys'}})
=
\frac{|O_{\Phi}(q)|^2}{(\alpha_A\alpha_B)^2}.
\label{eq:hyb_obs_def}
\end{equation}
The signed square root then recovers $O_{\Phi}(q)$. For data with finite-shots, this square-root estimator is consistent but can be biased because of the nonlinearity of the square root. 
The modulus-squared readout is therefore optional and is used only when a configuration-wise common sign has been established independently.

\section{Relation to Prior Work}\label{sec:related}
 
Several distinct programs seek quantum acceleration of Monte Carlo
methods. Having presented both the full-quantum and hybrid protocols,
we now delineate precisely how DQ$^{2}$MC differs from each, organized
by the axes that matter operationally and summarized in
Table~\ref{tab:comparison}: which object is placed on the quantum
processor, how the fermions are encoded, which classical bottleneck is
targeted, and what overhead is paid. We discuss each family in turn below:
\begin{enumerate}
\item  \emph{Quantum-assisted QMC (QC-AFQMC).} In the approach of Huggins
\textit{et al.}~\rcite{Huggins2022NatureQCQMC}, the Monte Carlo
algorithm, including the random walkers, importance sampling, and the probability
distribution itself, remains entirely classical; the quantum
processor supplies a \emph{trial wavefunction} beyond classical reach,
whose overlaps with walker states reduce the constraint bias of
phase-free AFQMC. The quantum resource therefore targets the bias of
the classical trial state, not the cost of sampling, and the required overlap or local-energy measurements can incur exponential sample complexity for some trial/walker pairs~\rcite{MazzolaCarleo2022Exponential}. In DQ$^{2}$MC, the distribution itself is placed on the quantum register, the sampling (or its coherent replacement) is the quantum step, and no trial wavefunction enters at finite temperature.
 
\item \emph{Quantum acceleration of Monte Carlo.} 
Szegedy-type quantum walks can quadratically improve certain gap-dependent detection and hitting tasks when coherent access to a reversible Markov kernel is available~\rcite{Szegedy2004Walk}. Quantum-proposal MCMC instead keeps a classical Metropolis accept/reject chain and uses a quantum circuit to generate proposals~\rcite{Layden2023QeMCMC}. These approaches accelerate Markov primitives supplied as input. The full quantum protocol of DQ$^{2}$MC  instead synthesizes the auxiliary-field distribution from the action matrix and removes the Markov chain from observable averaging, so no mixing or autocorrelation time appears.
 
\item \emph{Hybrid quantum-classical Monte Carlo and impurity solvers.} 
The DQ$^{2}$MC hybrid protocol of Sec.~\ref{sec:compat} is naturally compared with prior quantum-in-the-loop schemes, from which it differs along two axes: \emph{what is decomposed}, and \emph{what the quantum subroutine computes}. Quantum-classical DMFT~\rcite{BauerWeckerMillisHastings2016QDMFT,Kreula2016QDMFT} uses a quantum impurity solver inside a classical self-consistency loop, and the quantum subroutine solves a genuinely quantum many-body subproblem. QC-AFQMC decomposes the \emph{estimator}, a classical walk that queries the quantum computer for wavefunction overlaps.  
In contrast, the hybrid DQ$^{2}$MC holds a constant-size active block of HS variables as qubits while the remainder are classical controls, and its quantum subroutine performs an \emph{exact conditional (heat-bath) sample} of that block. The update procedure is rejection-free and optimization-free.

\item \emph{Thermal-state and Gibbs samplers.} Quantum Metropolis sampling~\rcite{Temme2011Metropolis}, its quantum-quantum variant~\rcite{YungAspuru2012QQMetropolis}, 
quantum phase estimation-based Gibbs preparation~\rcite{PoulinWocjan2009Gibbs}, Lindbladian samplers~\rcite{ChenKastoryanoBrandaoGilyen2023ThermalPrep,ChenKastoryanoGilyen2024GibbsSampler} directly target the physical many-body Gibbs state $\rho\propto e^{-\beta \hat H}$ 
, a purification thereof, or the ground state~\rcite{DingLin2024Lindbladian}; QITE-type methods~\rcite{Motta2020QITE,McArdle2019VITE} instead implement imaginary-time evolution, which can be embedded in protocols for thermal-state preparation or thermal-average estimation. DQ$^2$MC differs from these approaches at the level of the object being prepared. It prepares a coherent encoding of the classical auxiliary-field measure instead of the thermal density matrix directly.
In contrast to direct fermionic Gibbs-state preparation, DQ$^{2}$MC does not encode the physical fermionic modes as qubits: it operates on the single-particle action matrix, requires only $O(\log N_{\tn{st}})$ system qubits, and avoids Jordan--Wigner or Bravyi--Kitaev mappings~\rcite{JordanWigner1928,BravyiKitaev2002Fermion}. Gibbs-state preparation also suffers from the auto-correlation and mixing-time issues, which are absent in full-quantum DQ$^{2}$MC at the price of condition number overhead.

\item \emph{Classical-to-quantum probability loaders.} One might
prepare $\ket{\mathrm{DQ}^{2}\mathrm{MC}}$ with a generic amplitude
loader. The Grover-Rudolph protocol~\rcite{GroverRudolph2002} loads
only distributions whose marginals are classically integrable in
polynomial time, 
where no known routes prepares the DQMC marginal distribution efficiently.
; QRAM-based
encoding~\rcite{GiovannettiLloydMaccone2008QRAM,KerenidisPrakash2017}
lifts that restriction but requires all $2^{N_{\tn{st}}}$ weights to
be precomputed and stored, i.e., a completed classical DQMC
enumeration. The determinant oracle $\mathcal{O}_{\det}$ instead
\textit{synthesizes} the amplitudes in place from the action-matrix
oracle $\mathcal{U}_{M}$ at circuit-depth polynomial in $N_{\tn{st}}$,
with no classical precomputation of any weight.
\end{enumerate}

The key distinction is that DQ$^{2}$MC places the auxiliary-field measure itself on the quantum processor; its hybrid realization samples conditional blocks by rejection-free heat-bath updates. The dominant overhead lies in the postselected loading step, whose success probability is set by the shape of the full or conditional DQMC distribution (Sec.~\ref{sec:loading}). Table~\ref{tab:comparison} summarizes the approaches discussed above, with DQ$^{2}$MC in the final rows.

\begin{table*}[t]
\caption{\label{tab:comparison}
{\bf Comparison of quantum approaches to accelerating (quantum) Monte
Carlo.} ``Quantum object'' denotes what is placed on the quantum
processor; ``Encoding'' the fermion-to-qubit mapping (2nd quant.
$\equiv$ Jordan-Wigner/Bravyi-Kitaev on the many-body Hilbert space;
1st quant. $\equiv$ single-particle action matrix; --- $\equiv$ not
fermionic); ``Bottleneck targeted'' the classical cost addressed;
``Update'' the sampling primitive where applicable; ``Dominant
overhead'' the price paid. Hybrid cost components are detailed in Table~\ref{tab:hyb_cost}. }
\scriptsize
\setlength{\tabcolsep}{2pt}
\begin{ruledtabular}
\begin{tabular}{lllll}
Approach & Quantum object & Encoding & Update / bottleneck & Dominant overhead \\
\hline
QC-AFQMC~\rcite{Huggins2022NatureQCQMC} & trial state, overlaps & 2nd quant. & classical walk& overlap/local-energy samples~\rcite{MazzolaCarleo2022Exponential} \\
Quantum Walk~\rcite{Szegedy2004Walk} & Markov kernel & --- & Grover's reflection & pre-encoding the Markov kernel 
\\
Quantum-proposal MCMC~\rcite{Layden2023QeMCMC} & proposal moves & spin/Ising & propose $+$ accept/reject & mixing time \\
Quantum-classical DMFT~\rcite{BauerWeckerMillisHastings2016QDMFT,Kreula2016QDMFT} & impurity solution & 2nd quant. & self-consistency loop & impurity solve $+$ loop \\
Quantum Metropolis / Gibbs~\rcite{Temme2011Metropolis,YungAspuru2012QQMetropolis,PoulinWocjan2009Gibbs,ChenKastoryanoBrandaoGilyen2023ThermalPrep} & $\rho\propto e^{-\beta \hat H}$ & 2nd quant. & thermal-state preparation & gap / mixing / overlap \\
Grover-Rudolph~\rcite{GroverRudolph2002} & probability amplitude & --- & amplitude loading & marginal distribution \\
\hline
DQ$^{2}$MC, full quantum & aux.-field measure & 1st quant. & coherent; no chain & \shortstack[l]{QSVT, global postselection} \\
DQ$^{2}$MC, hybrid & cond. block measure & 1st quant. & rejection-free heat-bath & \shortstack[l]{QSVT, postselection, autocorrelation}\\
\end{tabular}
\end{ruledtabular}
\end{table*}

\section{Extensions and Generalizations}\label{sec:extensions}
 
Having demonstrated the DQ$^{2}$MC algorithm and its compatibility with both the full-quantum and hybrid quantum-classical setups, we turn to generalizations and extensions of the present framework. We discuss four directions: the zero-temperature extension of DQ$^{2}$MC to the projective quantum Monte-Carlo (PQMC) setting; the quantum-algorithmic counterpart of the fermion sign problem; the applicability of higher-order Gauss-Hermite quadratures as a generalization of Ising HS fields; and, finally, why the auxiliary-field route is worth taking at all when direct quantum simulation of the Hamiltonian is the obvious alternative.
 
\subsection{Zero-temperature Extension: $\mathrm{PQ^2MC}$}\label{sec:pq2mc}
 
The framework of Secs.~\ref{sec:qalg}--\ref{sec:compat} extends to zero-temperature projective quantum Monte-Carlo (PQMC), yielding PQ$^{2}$MC. In standard PQMC~\cite{SugiyamaKoonin1986, Sorella1989, AssaadEvertz2008}, the ground state $|\Psi_{\Theta}\rangle=e^{-\Theta H}|\Psi_{T}\rangle$ is obtained by stochastic imaginary-time evolution of a Slater-determinant trial state $|\Psi_{T}\rangle=\prod_{n=1}^{N_{p}}R_{nj}c^{\dagger}_{j}|0\rangle$, where $c_j^{\dag}$ denotes the electron creation operator for site $j$, $R_{nj}$ denotes the  trial wavefunction for the eigenstate $n$, and $N_{p}$ denotes total number of electrons. The HS-decomposed weights take the form $p_{\Phi}\propto\det[L^{\dagger}B_{\Phi}(2\Theta,0)R]$, where $L,R$ are the trial wavefunctions at the two boundaries of the imaginary-time strip, and $B_{\Phi}(2\Theta,0)$ is the imaginary-time evolution operator from $\tau=0$ to $\tau=2\Theta$ conditioned on the configuration $\Phi$. Naively, two obstacles prevent direct application of the protocol of Sec.~\ref{sec:qalg}: the equal-time Green's function $G_{\Phi}(0)$ is a projector, so the identity $p_{\Phi}=\det G_{\Phi}^{-1}(0)$ underlying the protocol no longer holds, and projection onto $|\Psi_{T}\rangle$ is non-unitary.
 
Both obstacles stem from the Slater-determinant formulation of canonical PQMC, and both are resolved by recasting PQMC as a coherent-state path integral with boundary Grassmann variables $\bar{\chi},\chi$ that absorb the trial wavefunctions (see Appendix~\ref{app:pqmc_path_integral}). The result is an extended action matrix,
\begin{subequations}
\beq
\label{eq:outlook_M_pqmc_defM}
M_{\Phi}^{(0)}
&=&
\begin{pmatrix}
\mathbb{1} &  &  &  & R \\
-B_{1} & \mathbb{1} &  &  &  \\
 & \ddots & \ddots &  &  \\
 &  & -B_{L_{T}} & \mathbb{1} &  \\
 &  &  & L^{\dagger} & \mathbb{0}
\end{pmatrix},
\qquad
\\
\det M_{\Phi}^{(0)}
&=&
\det\!\left[L^{\dagger}B_{\Phi}(2\Theta,0)R\right].
\label{eq:outlook_M_pqmc}
\eeq
\end{subequations}
Equation~\eqref{eq:outlook_M_pqmc_defM} has the same block-bidiagonal structure as the finite-temperature action matrix of Eq.~\eqref{eq::def_action_matrix}, with boundary blocks $L^{\dagger},R$ encoding the trial wavefunction. For a trial wavefunction with $N_p$ particles, $L^{\dagger}$ and $R$ are $N_{p}\times N_{s}$ matrices ($N_s\equiv L^2$ in two-dimension), so $M_{\Phi}^{(0)}$ is square with dimension $(N_{\tn{st}}+N_{p})$. To make the dimension compatible with the qubit encoding (i.e. a multiple of $N_s$), $M_{\Phi}^{(0)}$ can be padded with an $(N_{s}-N_{p})$-dimensional identity block, which leaves $\det M_{\Phi}^{(0)}$ and $[M_{\Phi}^{(0)}]^{-1}$ unchanged. Block-encoding the boundary blocks costs $O(1)$ additional ancillas using linear combination of unitaries (LCU) and incurs no extra post-selection suppression, since $L^{\dagger}$ and $R$ are one-body projectors with singular values in $\{0,1\}$. With this padded action matrix in hand, the state-preparation protocol of Secs.~\ref{sec:qalg}--\ref{sec:compat} applies directly.
 
Observables are evaluated in similar fashion. Denoting the imaginary-time evolution operator from $\tau$ to $\tau'$ by $B(\tau',\tau)$, and defining the boundary-anchored propagators $B_{<}(\tau)\equiv B(\tau,0)$ and $B_{>}(\tau)\equiv B(\beta_{\mathrm{eff}},\tau)$ with $\beta_{\mathrm{eff}}\equiv L_{T}\Delta\tau$, the time-ordered Green's function $G^{T}_{i\tau,j\tau'}\equiv{[M_{\Phi}^{(0)}]}^{-1}_{i\tau,j\tau'}$ takes the explicit form
\begin{subequations}
\label{eq:outlook_G_pqmc}
\beq
G^{T}_{i\tau,j\tau'}
&=&
\begin{cases}
[\mathbb{I}-\mathcal{P}_{\Phi}(\tau)]\,B(\tau,\tau'), & \tau\geq\tau',\\[4pt]
B(\tau',\tau)\,\mathcal{P}_{\Phi}(\tau), & \tau<\tau',
\end{cases}
\\
\mathcal{P}_{\Phi}(\tau)&\equiv& B_{<}(\tau)\,R\,{[L^{\dagger}B_{>}(0)R]}^{-1}L^{\dagger}\,B_{>}(\tau).
\eeq
\end{subequations}
Equation~\eqref{eq:outlook_G_pqmc} reduces to the standard PQMC propagator after shifting the imaginary-time origin via $\beta_{\mathrm{eff}}\to 2\Theta$, $\tau\to\Theta+\tau$. The equal-time block ($\tau=\tau'$) is a projector, but the full time-ordered $G^{T}$ remains invertible, so the QSVT-inverse construction of Sec.~\ref{sec:qalg_obs} applies without modification.
 
\subsection{Fermion Sign Problem}\label{sec:sign}
 
All constructions of Secs.~\ref{sec:qalg}--\ref{sec:compat} assumed a sign-problem-free HS decomposition, in which the two flavor contributions to the MC weights form complex-conjugate pairs, so that $p_{\Phi}=|\det M_{\Phi}|^{2}$ is non-negative. When this assumption fails, $\det M_{\Phi}=s_{\Phi}|\det M_{\Phi}|$ acquires a configuration-dependent phase $s_{\Phi}\in U(1)$, and observables must be reweighted in the standard way~\cite{LohGubernatis1990sign, TroyerWiese2005},
\begin{equation}
\langle O\rangle
=
\frac{\langle O\,s\rangle_{|p|}}{\langle s\rangle_{|p|}},
\qquad
\langle s\rangle_{|p|}\sim e^{-\beta V\Delta},
\label{eq:outlook_reweight}
\end{equation}
where $\langle\cdot\rangle_{|p|}$ denotes the average over the phase-quenched distribution $|\det M_{\Phi}|^{2}$ and $\Delta>0$ is the free-energy density difference between the phase-carrying and phase-quenched ensembles. Since the space-time volume obeys $\beta V=N_{\tn{st}}\Delta\tau$, it is convenient to define the free-energy gap per space-time site, $\tilde\Delta\equiv(\Delta\tau)\,\Delta$, so that $\langle s\rangle_{|p|}\sim e^{-N_{\tn{st}}\tilde\Delta}$. Resolving $\langle O\rangle$ to fixed additive precision then requires $O(\langle s\rangle_{|p|}^{-2})=O(e^{2N_{\tn{st}}\tilde\Delta})$ classical samples, which corresponds to the standard signal-to-noise catastrophe.
 
The DQ$^{2}$MC framework casts this catastrophe in transparent operational form. Applied to a sign-problematic model, the full quantum protocol of Sec.~\ref{sec:qalg} prepares the \emph{phase-quenched} state
\begin{equation}
|\Psi_{|p|}\rangle
\equiv
\frac{1}{\sqrt{\mathcal{N}_{|p|}}}
\sum_{\Phi}|\det M_{\Phi}|\,|\Phi\rangle,
\quad
\mathcal{N}_{|p|}\equiv\sum_{\Phi}|\det M_{\Phi}|^{2},
\label{eq:outlook_phaseless}
\end{equation}
where $|\Phi\rangle$ denotes, as throughout, an auxiliary-field basis state. Augmenting the same circuit with a phase oracle that imprints $s_{\Phi}$ would produce the \emph{phase-carrying} state
\begin{equation}
|\Psi_{s}\rangle
\equiv
\frac{1}{\sqrt{\mathcal{N}_{|p|}}}
\sum_{\Phi}s_{\Phi}|\det M_{\Phi}|\,|\Phi\rangle.
\label{eq:outlook_signed}
\end{equation}
The present construction does not provide such a phase oracle: the construction in Appendix~\ref{app:state_prep_oracle} acts on the \emph{singular values} of $M_{\Phi}$ through $(M_{\Phi}^{\dagger}M_{\Phi})^{1/2}$, which by definition discards the phase of $\det M_{\Phi}$. The correspondence below therefore holds \emph{given} such an oracle.
 
With both states in hand, and for an observable $O$ controlled by the auxiliary qubits, Eq.~\eqref{eq:outlook_reweight} is precisely the quantum \textit{weak value} of $O$,
\begin{equation}
\langle O\rangle
=
\frac{\langle\Psi_{|p|}|O|\Psi_{s}\rangle}{\langle\Psi_{|p|}|\Psi_{s}\rangle},
\label{eq:outlook_weakvalue}
\end{equation}
with pre-selection on the phase-carrying state $|\Psi_{s}\rangle$ and post-selection on the phase-quenched state $|\Psi_{|p|}\rangle$~\cite{AharonovAlbertVaidman1988, DresselRMP2014}. The DQ$^{2}$MC formalism therefore provides an \textit{exact} map between the fermion sign problem and weak-value extraction. Standard estimators of Eq.~\eqref{eq:outlook_weakvalue} are post-selection on $|\Psi_{|p|}\rangle$ or a SWAP test against $|\Psi_{s}\rangle$, which pay a shot overhead $O(|\langle\Psi_{|p|}|\Psi_{s}\rangle|^{-2})\sim O(e^{2N_{\tn{st}}\tilde\Delta})$. This exponential overhead is in precise correspondence with the classical reweighting cost above: the exponentially small overlap between the two states \emph{are} the exponentially small average sign. Several weak-value protocols have been proposed in the quantum-information literature~\cite{KedemVaidman2010, MitchisonJozsaPopescu2007, ReschSteinberg2004, Dressel2015interference, Wagner2024weakvalue}, but none is presently known to remove the exponential overhead in the regime $N_{\tn{st}}\tilde\Delta\gg 1$ relevant here.

\subsection{Higher-order Gauss-Hermite Quadrature}\label{sec:higher_GH}
 
The formulation of DQ$^{2}$MC based on Ising-type HS decoupling with binary HS variables generalizes naturally to higher-order Gauss-Hermite quadratures. The decoupling adopted in Eq.~\eqref{eq::hs} is a 2-node Gauss-Hermite quadrature of the local interaction vertex, incurring a $O(\Delta\tau)$ error per imaginary-time step~\cite{Hirsch1983discrete}. Consider instead the 4-node quadrature~\cite{MotomeImada1997},
\begin{equation}
e^{\Delta\tau\lambda\hat{A}^{2}}
\simeq
\frac{1}{4}\sum_{l=\pm 1,\pm 2}\gamma(l)\,e^{\sqrt{\Delta\tau\lambda}\,\eta(l)\,\hat{A}},
\label{eq:outlook_GH4}
\end{equation}
with weights $\gamma(\pm 1)=1+\sqrt{6}/3$, $\gamma(\pm 2)=1-\sqrt{6}/3$ and nodes $\eta(\pm 1)=\pm\sqrt{2(3-\sqrt{6})}$, $\eta(\pm 2)=\pm\sqrt{2(3+\sqrt{6})}$. Equation~\eqref{eq:outlook_GH4} reduces the quadrature error to $O(\Delta\tau^{2})$ at the cost of promoting the binary HS variable $\Phi_{r,\tau}\in\{\pm 1\}$ to a four-valued one, $l\in\{\pm 1,\pm 2\}$. To implement this on top of the $\mathrm{DQ^2 MC}$ frame work, the configuration-dependent quadrature weights $\Gamma[\Phi]$ retained in Eq.~\eqref{eq::Z_dqmc} are no longer constant and must be carried through the magnitude; we thus introduce an additional magnitude ancilla to take this into account as follow.
 
The auxiliary register at each space-time point splits into a magnitude qubit $|\gamma\rangle$ and a sign qubit $|\eta\rangle$: the magnitude qubit selects between the two values of $|\gamma|$, while the sign qubit selects the sign of $\eta$. The HS-controlled one-body vertex $U_{\gamma,\eta}(\Phi)$ is conditioned on both qubits, replacing the Ising vertex inside the single-slice propagator $B_{\tau}$ of Sec.~\ref{sec:overview-dqmc}. Instead of a uniform superposition, the local auxiliary register begins in the amplitude-weighted state $\sum_{l}\sqrt{\gamma(l)/4}\,|l\rangle$, prepared by a fixed single-qubit rotation $G|0\rangle_{\gamma}\equiv\sqrt{\gamma(1)/2}\,|0\rangle+\sqrt{\gamma(2)/2}\,|1\rangle$ on the magnitude qubit together with a Hadamard on the sign qubit. This is precisely the amplitude-weighted generalization of the uniform initialization of Sec.~\ref{sec:loading}. Beyond doubling the auxiliary register from $N_{\tn{st}}$ to $2N_{\tn{st}}$ qubits, the rest of the pipeline is unchanged. The construction extends straightforwardly to $n$-node quadratures at the cost of a $\lceil\log_{2}n\rceil$-qubit local register per space-time point.
 
\subsection{Why Auxiliary Fields?}\label{sec:why_aux}

We close by addressing a natural question: given a fault-tolerant quantum computer with many thousands of qubits, why route finite-temperature fermion simulation through auxiliary fields at all, rather than simulating the Hamiltonian directly?
 
The answer is that direct simulation does not remove the difficulty; it relocates it. A quantum computer natively implements unitary dynamics $e^{-i\hat H t}$, but the thermal expectation values that DQMC targets require preparing $\rho\propto e^{-\beta \hat H}$, for which no unitary suffices. Most known routes, e.g. quantum Metropolis, Davies-Lindblad samplers, or QITE, carry a runtime governed by a mixing time or spectral gap that is not known to be polynomial for interacting fermions in the strongly correlated regime, and the hardness of thermalization in generic many-body systems has been made precise~\cite{Temme2011Metropolis,ChenKastoryanoBrandaoGilyen2023ThermalPrep,ChenHuangPreskillZhou2024NaturePhys}. The post-selection overhead of Sec.~\ref{sec:loading} and the mixing time of a direct thermal sampler are thus two prices for the same underlying difficulty; which one is smaller is model- and regime-dependent.
 
What the auxiliary-field route buys is the \emph{form} in which that difficulty appears. In DQ$^{2}$MC the obstruction is the condition number of an explicit classical probability distribution: it can be measured in trial runs (Appendix~\ref{app:hubbard_preconditioning}), preconditioned by a single scalar (Sec.~\ref{sec:loading}), or tackled with the full classical Monte-Carlo toolbox through the hybrid protocol of Sec.~\ref{sec:compat}. No comparable handle is available for the mixing time of a Lindbladian sampler. Three further consequences follow. First, resources: the construction operates on the $N_{\tn{st}}$-dimensional single-particle action matrix, requiring $O(\log N_{\tn{st}})$ system qubits and no Jordan-Wigner or Bravyi-Kitaev encoding~\cite{JordanWigner1928,BravyiKitaev2002Fermion}, whereas second-quantized simulation demands a qubit per spin-orbital together with the associated string overhead. Second, observables: Wick's theorem delivers momentum- and Matsubara-resolved correlators directly from $N_{\tn{st}}$-dimensional matrices (Sec.~\ref{sec:qalg_obs}), whereas extracting time-displaced correlators from a prepared Gibbs state requires additional interferometric machinery. Third, verifiability: the protocol inherits the error structure of Monte Carlo such as detailed balance [Eq.~\eqref{eq:compat_db}], statistical error bars, and configuration-by-configuration agreement with classical DQMC in sign-free regimes, whereas certifying that a sampler has actually reached $e^{-\beta \hat H}/Z$, rather than having stopped mixing early, is itself hard.
 
None of this argues that direct simulation is the wrong long-term target; the two approaches are complementary. But for the near and intermediate term, an algorithm whose bottleneck is a diagnosable classical quantity, whose system register is logarithmically small, and whose hybrid form runs on a constant-size quantum block with classical inter-node communication, occupies a different and more immediately accessible point in the design space.

\section{Conclusion}\label{sec:conclusion}
 
In this work we have introduced determinant quantum-quantum Monte Carlo (DQ$^{2}$MC), a quantum algorithm that lifts classical DQMC onto coherent quantum registers. The central primitive is a determinant oracle $\mathcal{O}_{\det}$ that \textit{synthesizes} the DQMC amplitudes $\sqrt{p_{\Phi}}$ from a block encoding of the action matrix together with two polynomial QSVT chains, rather than loading them from classically precomputed data. Built on this primitive, we constructed two protocols and analyzed the sampling and observable-estimation procedures for each.
 
The full-quantum protocol holds the entire auxiliary-field register coherently, so that observable averages are interference amplitudes and no Markov chain, and hence no autocorrelation time, exists; it uses $O(N_{\tn{st}})$ auxiliary qubits. The hybrid quantum-classical protocol retains only a constant-size active block of qubits, reducing the system register to $O(\log N_{\tn{st}})$ while replacing the Metropolis-Hastings acceptance step with an exact heat-bath draw, so that cluster updates of any size are rejection-free and cost no more per update than single-site moves. Because only classical information passes between updates, the costs of successive updates add rather than multiply, classical acceleration techniques such as parallel tempering apply without modification, and the protocol admits distributed execution across multiple quantum processing units with purely classical communication. Both protocols operate entirely at the level of the single-particle action matrix, without Jordan-Wigner or Bravyi-Kitaev string operators, and both achieve a per-oracle-call circuit-depth scaling more favorably with the spatial volume $N_s$ in the optimal case than classical DQMC, at the price of a polynomial overhead in the number of Trotter steps and a post-selection overhead. Both agree quantitatively with exact DQMC on the half-filled square-lattice Hubbard model in a noise-free benchmark.
 
That post-selection overhead is the principal limitation, and we have pinpointed its origin. At the optimal preconditioning it is determined exactly by the largest target probability, $p_{\mathrm{succ}}=1/(2^{N_{\tn{st}}}p_{\max})$, which is polynomial when the largest weight lies within a polynomial factor of the uniform weight and exponentially small for sharply peaked distributions (Sec.~\ref{sec:loading}). We isolated its origin to the probability-loading step and related it to the difficulty of preparing thermal pure quantum states in generic quantum imaginary-time evolution (Sec.~\ref{sec:qalg_summary}).  We regard the sharp form of this criterion as itself a useful output: it converts a vague question about quantum hardness into a checkable property of an explicit classical distribution.

The same coherent state-preparation primitive supports several extensions. We described a zero-temperature counterpart of projector quantum Monte Carlo, PQ$^{2}$MC, built from an extended action matrix with boundary blocks encoding the trial wavefunction (Sec.~\ref{sec:pq2mc}); and the accommodation of higher-order Gauss-Hermite quadratures, which reduce the Trotter error at the cost of an enlarged auxiliary register (Sec.~\ref{sec:higher_GH}).
 
Finally, the framework yields an exact correspondence between the fermion sign problem and quantum measurement theory. The standard reweighting estimator (a phase-quenched ensemble corrected by the fermion sign) is precisely a quantum weak value, with pre-selection on the sign-carrying state and post-selection on the phase-quenched one, and the exponentially rare post-selection events of weak-value extraction \emph{are} the exponentially small average sign (Sec.~\ref{sec:sign}). This does not solve the sign problem; if anything it suggests why the problem should remain hard on quantum hardware, at least under the frame of $\mathrm{DQ^2 MC}$. 
The weak-value and classical-reweighting formulations therefore carry the same exponential measurement overhead.
The weak-value formulation provides an operational interpretation of the exponential reweighting cost and identifies a concrete connection between the fermion sign problem and quantum measurement theory. 

Several open questions follow from the present work. On state preparation in the context of the full quantum protocol, a classification of Hamiltonians whose DQMC distributions meet the polynomial success-probability criterion remains an interesting question for future study. 
A warm-start may also mitigate this success-probability issue via replacing the uniform initial state by $|\psi_{q}\rangle=\sum_{\Phi}\sqrt{q_{\Phi}}\ket{\Phi}$ for some classically tractable approximation $q_{\Phi}\simeq p_{\Phi}$.
In this case, the success probability is governed by the Bhattacharyya-type overlap ${\big(\sum_{\Phi}\sqrt{p_{\Phi}q_{\Phi}}\big)}^{2}$ between the quantum target and the initial state. Natural choices of $q_{\Phi}$ include a product-state mean-field marginal, a Gaussian-fluctuation approximation around a Hartree-Fock saddle \cite{ZhangKrakauer2003}, or even a trial run of the classical DQMC itself to generate a sample-based approximation. 
Warm starting is compatible with the rest of the pipeline without modification. 
An adiabatic-annealing route from $|\psi_q\rangle$ to $|\psi_p\rangle$~\cite{KadowakiNishimori1998,AlbashLidar2018RMP} is unlikely to remove the bottleneck: each annealing step is prepared probablistically and thus the total $p_{\tn{succ}}$ is a product of $p_{\tn{succ}}$ for each step.
Whether a different bounded state-loading transformation, including an alternative non-exponential target in the clipped regime, can achieve $o\!\left(\sqrt{N_{\tn{st}}}\right)$ exp-QSVT degree over the full faithful interval with fewer hyperparameters, and whether a coherent low-rank update (e.g. a quantum version of Sherman-Morrison technique) can reduce the $\widetilde O(N_{\tn{st}}^{3/2})$ determinant-oracle depth paid by each hybrid block update remains an interesting question. Constructing a phase-sensitive determinant oracle involving a Hermitian dilation or eigenvalue transformation yielding $\log\det M_{\Phi}$ rather than $\log|\det M_{\Phi}|$, or phase estimation on a suitably constructed unitary~\cite{Kitaev1995QPE} remain interesting possibilities. Given such an oracle, any weak-value protocol that removed the exponential overhead would translate through Eq.~\eqref{eq:outlook_weakvalue} into progress on sign-problematic DQMC, modulo the state-preparation overhead of Sec.~\ref{sec:loading}; whether this map can be exploited or sharpened into a hardness equivalence remains to be investigated. Whether intrinsic noise on quantum hardware can be utilized to improve the protocols introduced here is also an interesting open direction.

\section{Acknowledgments} 
We thank D. Pimenov for helpful discussions.
This work is supported in part by a grant from the Department of Energy (DE-SC0026112) under the Early Career Research Program to DC.
XW acknowledges the use of large language models (ChatGPT 5.5 and Claude Opus 4.6) for research assistance.

\appendix
\addtocontents{toc}{\protect\hideappendixsubsectionsintoc}

\section{State-Preparation Oracle and Gate-Level Construction}\label{app:state_prep_oracle}

\begin{figure*}[t]
\centering
\includegraphics[width=15cm]{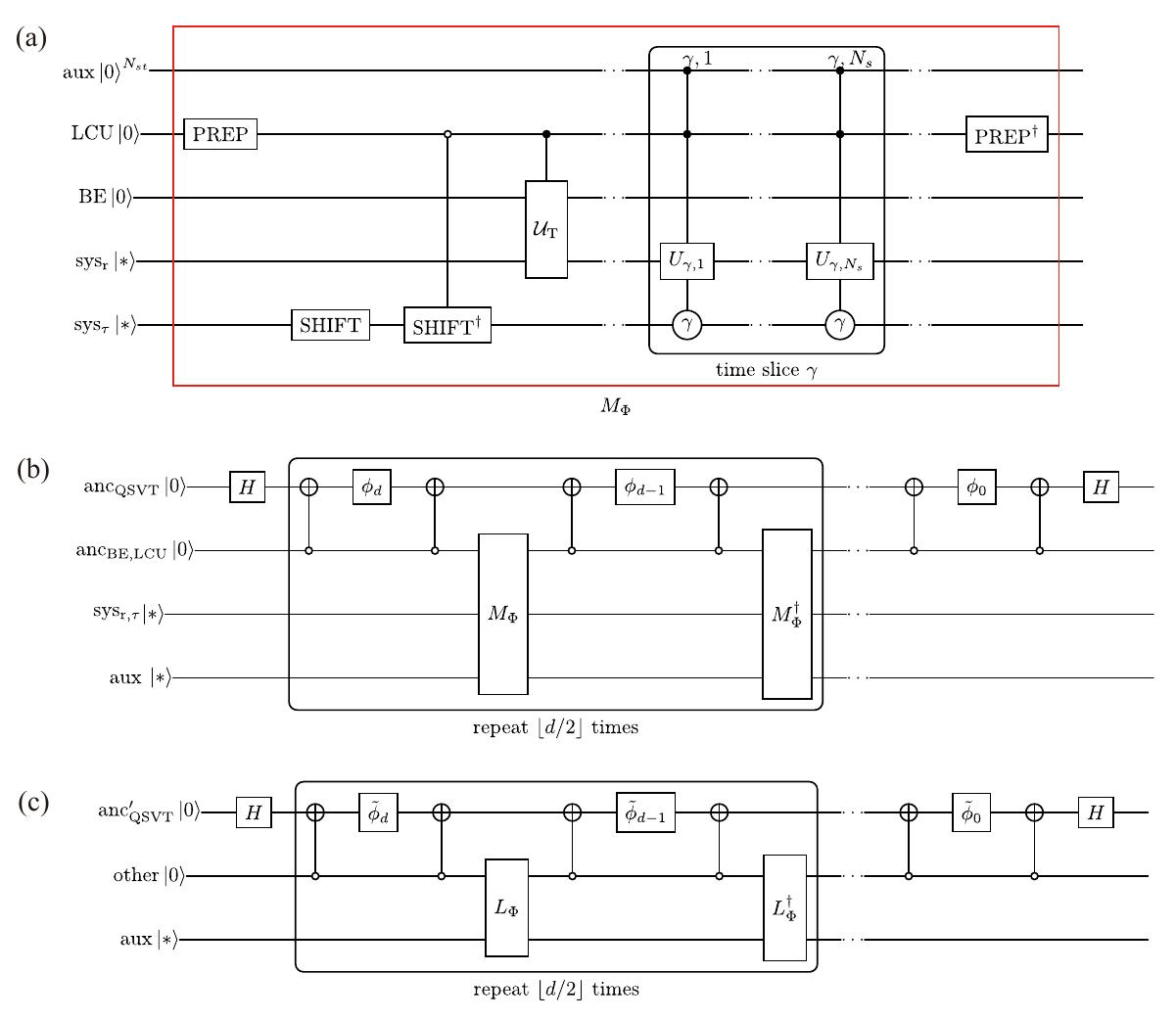}
\caption{\label{Fig::fullq_gates} {\bf Gate-level construction of the determinant oracle.} (a) Action-matrix oracle $\mathcal{U}_{M}$, block-encoding $M_{\Phi}/\alpha_M$: a single-qubit LCU (PREP, PREP$^{\dagger}$) selects between the cyclic shift $U_{\mathrm{shift}}^{\dagger}$ on the time-slice register $\mathtt{sys}_{\tau}$ (where circle denotes select-gate) and the kinetic-propagator block encoding $\mathcal{U}_{T}$ dressed by the HS-controlled one-body vertices $U_{\gamma,i}$ on the spatial register $\mathtt{sys}_{r}$, with the $\mathtt{aux}$ register selecting the HS configuration. The red box encloses the preconditioned action matrix $ M_{\Phi}$ in Eq.~\eqref{eq:qalg_Mtilde}. (b) QSVT-log sequence $\mathcal{U}_{\log}$: an even-parity QSVT polynomial approximating $\log|x|$ is realized by interleaving phases $\phi_{0},\ldots,\phi_{d}$ with $\lfloor d/2\rfloor$ alternating calls to $M_{\Phi}$ and $M_{\Phi}^{\dagger}$ (with $\mathtt{anc}_{\mathrm{BE},\mathrm{LCU}}$ collecting the BE and LCU ancillas of panel (a)); combined with a maximally entangled system-copy pair via the trace trick of Eq.~\eqref{eq:qalg_trace} it yields the scalar log-trace block $L_{\Phi}$ of Eq.~\eqref{eq:qalg_logsv}. (c) exp-QSVT sequence $\mathcal{U}_{\exp}$: a QSVT polynomial approximating the exponential window $f_{\exp}$ of Eq.~\eqref{eq:qalg_expwindow} is applied to $L_{\Phi}$ via phases $\tilde\phi_{0},\ldots,\tilde\phi_{d}$ and alternating calls to $L_{\Phi}, L_{\Phi}^{\dagger}$; postselecting the work registers $\mathtt{anc}=\{\mathtt{anc}'_{\mathrm{QSVT}},\mathtt{other}\}$ on $|0\rangle$ implements the determinant oracle on the $\mathtt{aux}$ register.}
\end{figure*}

In this appendix, we elaborate further on the details behind the construction pipeline in Table~\ref{tab:state_prep_interface} for the determinant oracle $\mathcal{O}_{\det} = \mathcal{U}_{\exp} \circ L_{\Phi} \circ \mathcal{U}_{\log} \circ \mathcal{U}_M$. The key is to first construct a block encoding of the action matrix $M_{\Phi}$, then to use QSVT and coherent trace estimation to realize the identity, 
\begin{equation}
\log\left|\det M_{\Phi}\right|
=
\operatorname{Tr}\log\sqrt{M_{\Phi}^{\dagger}M_{\Phi}},
\label{eq:qalg_logdet}
\end{equation}
followed by exponentiation to yield the diagonal determinant oracle. The construction is summarized visually in Fig.~\ref{Fig::fullq_gates}.

\subsection{Action-Matrix Block Encoding $\mathcal{U}_{M}$}\label{app:oracle_um}

The first step is to construct an oracle for the action matrix $M_{\Phi}$ itself on the system register. A gate-level construction for the action matrix is shown in Fig.~\ref{Fig::fullq_gates}(a).  It is useful to isolate the purely cyclic shift along the Trotter direction. Let $U_{\mathrm{shift}}$ denote the operator acting on the time-slice register, and let $\mathcal{V}_{\Phi}$ denote the full HS-controlled interaction vertex, defined as
\begin{equation}
\mathcal{V}_{\Phi}\equiv
\operatorname{diag}\!\left(
-U_{1}(\Phi),\,
\ldots,\,
-U_{L_{\mathrm{T}}-1}(\Phi),\,
U_{L_{\mathrm{T}}}(\Phi)
\right),
\label{eq:qalg_Vphi}
\end{equation}
where $U_{\tau}(\Phi)$ is the one-body interaction vertex on the imaginary-time slice $\tau$ defined as the product of one-body vertices on each spatial site,
\begin{equation}
U_{\tau}(\Phi)=\prod_{i}U_{\tau,i}(\Phi).
\label{eq:qalg_Upiece}
\end{equation}
Here, $U_{\tau,i}(\Phi)$ is the unitary one-body vertex obtained after the HS decoupling conditioned on the auxiliary field $\Phi$. To match the circuit notation, we write $U_{\gamma,i}\equiv U_{\tau,i}(\Phi)|_{\tau=\gamma}$ when $\gamma$ labels the selected time slice.
With these definitions, the fermion action matrix can be rewritten as
\begin{equation}
\begin{aligned}
M_{\Phi}&=U_{\mathrm{shift}}\widetilde M_{\Phi},\\
\widetilde M_{\Phi}
&\equiv U_{\mathrm{shift}}^{\dagger}
+\left(\mathbb{1}_{\tau}\otimes e^{-\Delta\tau T}\right)\mathcal{V}_{\Phi}.
\end{aligned}
\label{eq:qalg_Mtilde}
\end{equation}
The advantage of Eq.~\eqref{eq:qalg_Mtilde} is that all HS dependence resides in controlled one-body unitaries.
The only nonunitary object is the kinetic propagator $e^{-\Delta\tau T}$, which can be block-encoded separately by introducing a BE ancilla register (denoted as $\mathtt{BE}$ in Fig.~\ref{Fig::fullq_gates}a).
Choosing a normalization $\alpha_T$ such that all the singular values of $e^{-\Delta\tau T}/\alpha_T$ lie in $[0,1]$, we construct a one-ancilla block encoding
\begin{equation}
\mathcal{U}_{T}
\;:\;
\left(\langle 0|_{\mathtt{BE}}\otimes \mathbb{1}_{\mathtt{sys}}\right)\mathcal{U}_{T}
\left(|0\rangle_{\mathtt{BE}}\otimes \mathbb{1}_{\mathtt{sys}}\right)
=
\frac{e^{-\Delta\tau T}}{\alpha_T}.
\label{eq:qalg_BE_T}
\end{equation}
The two terms in Eq.~\eqref{eq:qalg_Mtilde}, the shift operator $U_{\mathrm{shift}}^{\dagger}$ and the kinetic contribution $e^{-\Delta\tau T}$ dressed by the HS-controlled vertices $\mathcal{V}_{\Phi}$, are combined via a linear combination of unitaries (LCU) with one ancilla qubit (denoted as $\mathtt{LCU}$). The resulting fermion action-matrix oracle $\mathcal{U}_{M}$ acts coherently on the system and ancilla registers and satisfies, for each computational-basis auxiliary configuration,
\begin{equation}
\left(
\langle 0|_{\mathtt{LCU}}
\langle 0|_{\mathtt{BE}}
\otimes
\mathbb{1}
\right)
\mathcal{U}_{M}(\Phi)
\left(
|0\rangle_{\mathtt{LCU}}
|0\rangle_{\mathtt{BE}}
\otimes
\mathbb{1}
\right)
=
\frac{M_{\Phi}}{\alpha_M},
\label{eq:qalg_BE_M}
\end{equation}
with a \textit{fixed} normalization factor $\alpha_M = 1+\alpha_T$. The explicit gate-level construction of the fermion action-matrix oracle $\mathcal{U}_{M}$ is shown in Fig.~\ref{Fig::fullq_gates}(a).

\subsection{QSVT of Logarithm $\mathcal{U}_{\log}$ (QSVT-log)}\label{app:oracle_ulog}

The second step is to implement Eq.~\eqref{eq:qalg_logdet}, starting by  applying an even-parity quantum singular value transformation (QSVT) polynomial approximating $\log |x|$ on the singular-value interval of $M_{\Phi}/\alpha_M$.

Specifically, we obtain the QSVT phases $\phi_{0},\ldots,\phi_{d}$ from the Chebyshev polynomial expansion $f_{\tn{approx}}(x)$ of $\log |x|$ on the interval $|x|\in[\kappa^{-1}_{\log},1]$.
Here $\kappa_{\log}$ serves as a global hyperparameter controlling the tolerance of the probability fidelity, analyzed in Sec.~\ref{sec:loading}.
We denote the full gate set (in Fig.~\ref{Fig::fullq_gates}(b)) realizing QSVT-log as $\mathcal{U_{\mathrm{log}}}$,  defined as
\begin{equation}
\mathcal{U_{\mathrm{log}}}
:
\left(\langle 0|_{\mathtt{log}}\otimes \mathbb{1}\right)\mathcal{U_{\mathrm{log}}}
\left(|0\rangle_{\mathtt{log}}\otimes \mathbb{1}\right)
=
\frac{1}{\alpha_{\log}}\log\bigg[\sqrt{M_{\Phi}^{\dagger}M_{\Phi}}\bigg],
\label{eq:qalg_logsv0}
\end{equation}
up to a field-independent additive constant  associated with the normalization $\alpha_M$. In Eq.~\eqref{eq:qalg_logsv0}, the ancilla quantum register $\mathtt{log}$ denotes the collection $\mathtt{log}\equiv\{\mathtt{BE}, ~\mathtt{LCU}, ~\mathtt{anc_{\mathtt{QSVT}}}\}$, and $\mathbb{1}$ acts on the $\mathtt{sys}$ and $\mathtt{aux}$ registers.

Since the QSVT block must have norm at most unity while $|\log|x||$ reaches $\log\kappa_{\log}$ on this interval, the polynomial approximates the \emph{rescaled} logarithm $\log|x|/\alpha_{\log}$ with a fixed normalization $\alpha_{\log}\gtrsim\log\kappa_{\log}$, where $\alpha_{\log}$ is an intrinsic $O(1)$ constant.
One further requirement on the tolerance deserves emphasis. 
Imposing the global tolerance $|f_{\tn{approx}}(x)-\log|x||<\epsilon_{\log}$ per singular value produces an additive error of at most $\epsilon_{\log}$, the depth scaling of the QSVT-log is $O(\kappa_{\log}\log\epsilon^{-1}_{\log})$ \cite{GilyenSuLowWiebe2019QSVT, LowChuang2019qubitization}. As we will further show in Appendix.~\ref{app:oracle_uexp}, faithful preparation of the DQMC weights to relative error $\epsilon$ therefore requires $\epsilon_{\log}\lesssim\epsilon/N_{\tn{st}}$, so the query complexity of the log-QSVT operation is $O\!\left(\kappa_{\log}\log (N_{\tn{st}}/\epsilon)\right)$; the space-time volume enters only logarithmically.

\subsection{Coherent Trace Estimation $L_{\Phi}$}\label{app:oracle_lphi}

The third step is to evaluate the trace in Eq.~\eqref{eq:qalg_logdet} coherently, which requires an ancillary copy of the system register, denoted ${\mathtt{sys'}}$. We start with a maximally entangled pair between the system registers ${\mathtt{sys}}$ and ${\mathtt{sys'}}$, prepared unitarily by a transversal layer of Hadamard and CNOT gates. For any matrix $A$ acting on the $N_{\tn{st}}$-dimensional system register,
\begin{equation}
\begin{aligned}
\frac{1}{N_{\tn{st}}}\operatorname{Tr} A
&=
\left(
\frac{1}{\sqrt{N_{\tn{st}}}}\sum_i \langle i|_{\mathtt{sys'}}\langle i|_{\mathtt{sys}}
\right)
(\mathbb{1}\otimes A)
\\
&\quad\times
\left(
\frac{1}{\sqrt{N_{\tn{st}}}}\sum_j |j\rangle _{\mathtt{sys'}}|j\rangle_{\mathtt{sys}}
\right).
\end{aligned}
\label{eq:qalg_trace}
\end{equation}
Applying Eq.~\eqref{eq:qalg_trace} to the operator in Eq.~\eqref{eq:qalg_logsv0} therefore yields (up to the rescaling $\alpha_{\log}$) the configuration-dependent scalar
\begin{equation}
x_{\Phi}
\equiv
\frac{1}{N_{\tn{st}}}
\operatorname{Tr}
\log\frac{\sqrt{M_{\Phi}^{\dagger}M_{\Phi}}}{\alpha_M}
=
\frac{1}{N_{\tn{st}}}
\log\left|\det \frac{M_{\Phi}}{\alpha_M}\right|.
\label{eq:qalg_xphi}
\end{equation}
We denote the corresponding scalar block after taking the trace by $L_{\Phi}$, which corresponds to post-selecting on $|0\rangle$ for all the `other' registers, defined as $\mathtt{other}\equiv\{\mathtt{log},~ \mathtt{sys},~\mathtt{sys'}\}$, after the maximally entangled pair in Eq.~\eqref{eq:qalg_trace} is unprepared.
The gate block $L_{\Phi}$ is then a block encoding of the diagonal operator whose entries are the rescaled log-weights,
\begin{equation}
L_{\Phi}
:
\left(\langle 0|_{\mathtt{other}}\otimes \mathbb{1}\right)L_{\Phi}
\left(|0\rangle_{\mathtt{other}}\otimes \mathbb{1}\right)
=
\sum_{\Phi} \frac{x_{\Phi}}{\alpha_{\log}} |\Phi\rangle\langle\Phi|,
\label{eq:qalg_logsv}
\end{equation}
which is related to $\tn{diag}\{\log p_{\Phi}\}$ by a configuration-independent affine map. The $\mathbb{1}$ in Eq.~\eqref{eq:qalg_logsv} acts on the $\mathtt{aux}$ registers.
At this stage, the construction has produced coherent access to the configuration-resolved log determinant, or equivalently to the diagonal log-weight data in the auxiliary-field registers.

\subsection{QSVT of Exponentiation $\mathcal{U}_{\exp}$}\label{app:oracle_uexp}

The final step in the determinant-oracle construction is to exponentiate the scalar diagonal block $L_{\Phi}$. This is achieved by a second QSVT sequence whose target function, expressed in terms of the log-weight data $x_{\Phi}$, is a bounded exponential window
\begin{equation}
f_{\exp}(x_{\Phi})\simeq 
\tn{min}\{1,\exp\!\left[N_{\tn{st}}(x_{\Phi}+x_0)\right]\},
\label{eq:qalg_expwindow}
\end{equation}
over the range of $x_{\Phi}$ retained by the log-QSVT approximation.
Here, $x_0$ is a field-independent preconditioning shift whose specific choice is discussed in Sec.~\ref{sec:loading}. The new QSVT ancilla is denoted by $\mathtt{anc}'_{\mathrm{QSVT}}$, while $\mathtt{other}$ collects the postselected registers of the log-trace block. For compactness, let
\begin{equation}
|0\rangle_{\mathtt{anc}'_{\mathrm{QSVT}}}
|0\rangle_{\mathtt{other}}
=
|0\rangle_{\mathtt{anc}} .
\end{equation}
The corresponding gate set of the exp-QSVT is shown in Fig.~\ref{Fig::fullq_gates}(c), and is denoted as $\mathcal{U}_{\exp}$.
The exp-QSVT gate set $\mathcal{U}_{\exp}$ is then a block encoding satisfying
\begin{equation}
\begin{aligned}
\mathcal{U}_{\exp}
:
&\left(\langle0|_{\mathtt{anc}}\otimes \mathbb{1}_{\mathtt{aux}}\right)
\mathcal{U}_{\exp}
\left(|0\rangle_{\mathtt{anc}}\otimes \mathbb{1}_{\mathtt{aux}}\right)
\\
&\qquad
=
\sum_{\Phi} f_{\exp}(x_{\Phi})
|\Phi\rangle\langle\Phi|_{\mathtt{aux}} ,
\end{aligned}
\label{eq:qalg_exp_block}
\end{equation}
which, together with the action-matrix oracle, the log-QSVT, and the coherent trace estimation above, realizes the determinant oracle $\mathcal{O}_{\det}$ in Eq.~\eqref{eq::det_oracle} up to the approximation error of the QSVT steps. The depth scaling for QSVT-exp is $O\!\left(\sqrt{N_{\tn{st}}\,\log(1/\varepsilon_{\exp})}\right)$, with a detailed derivation in Appendix~\ref{app:exp_qsvt_degree}.

Let us end by discussing the effect of the intermediate tolerance $\epsilon_{\log}$ in QSVT-log on the final prepared distribution. In the (averaged) log-weight $x_{\Phi}$ of Eq.~\eqref{eq:qalg_xphi}, the trace averages the per-value errors rather than accumulating them, but the subsequent exponentiation in Eq.~\eqref{eq:qalg_expwindow} amplifies this to a \emph{multiplicative} weight error $e^{N_{\tn{st}}\epsilon_{\log}}$. Faithful preparation of the DQMC weights to relative error $\epsilon$ therefore requires $\epsilon_{\log}\lesssim\epsilon/N_{\tn{st}}$, so the query complexity of the log-QSVT operation is $O\!\left(\kappa_{\log}\log (N_{\tn{st}}/\epsilon)\right)$ \cite{GilyenSuLowWiebe2019QSVT, LowChuang2019qubitization}; the space-time volume enters only logarithmically.
Combining the two QSVT sequence, the circuit depth scales as
$O\big(\kappa_{\tn{log}}\sqrt{\log\epsilon^{-1}_{\exp}}\log (N_{\tn{st}}/\epsilon)\, N_{\tn{st}}^{3/2}\big)$.

\section{Hubbard Model Benchmark}\label{app:hubbard_benchmarks}

In this appendix we present our benchmarking computations on the standard Hubbard model,
\begin{equation}
H=-t\sum_{\langle r,r'\rangle,\sigma}c_{r\sigma}^{\dagger}c_{r'\sigma}+U\sum_{r}\left(n_{r\uparrow}-\frac{1}{2}\right)\left(n_{r\downarrow}-\frac{1}{2}\right),
\end{equation}
where $t$ is the nearest-neighbor hopping amplitude and $U$ is the on-site repulsion. We will classically emulate the quantum algorithm on a $2\times 2$ cluster at half filling. Throughout we set $t=0.5$ and $U=1$ (i.e. $U/t=2$) and choose $\Delta\tau=0.2$ and $\beta=0.4$ for the Trotter discretization, corresponding to $L_{\mathrm{T}}=2$ time slices, $N_s=4$ sites, and hence $N_{\tn{st}}=L_{\mathrm{T}}N_s=8$ auxiliary fields spanning $2^{8}=256$ HS configurations, which is small enough that all circuit amplitudes can be evaluated exactly.

\subsection{Density-channel HS Decomposition}\label{app:hubbard_density_hs}
Here we show that under the density-channel HS decomposition, the model satisfies both assumptions of Sec.~\ref{sec:qalg}.
We write the repulsive interaction in terms of the centered density, $\delta n_{i\sigma}\equiv n_{i\sigma}-1/2$, and decouple it in the density channel. With the two-node Gaussian quadrature used in the simulations, the local vertex at site $i$ and time slice $\ell$ is
\begin{equation}
e^{-\Delta\tau U\delta n_{i\uparrow}\delta n_{i\downarrow}}
=
\frac{e^{\Delta\tau U/4}}{2}
\sum_{\Phi_{i,\ell}=\pm1}
e^{i\lambda\Phi_{i,\ell}\left(\delta n_{i\uparrow}+\delta n_{i\downarrow}\right)}
+O(\Delta\tau^2),
\label{eq:app_hubbard_hs_identity}
\end{equation}
where $\lambda=\sqrt{\Delta\tau U}$. Thus each spin flavor sees the same imaginary density vertex $\exp[i\lambda\Phi_{i,\ell}(n_{i\sigma}-1/2)]$, which is unitary and thus is satisfying Assumption 2 of Sec.~\ref{sec:qalg}. In this convention the scalar phase generated by the $-1/2$ shift is part of the single-spin weight. Writing $\Phi_{\ell}$ for the diagonal matrix with entries $\Phi_{i,\ell}$, the single-flavor weight is
\begin{equation}
 w_\uparrow[\Phi]
=
\exp\left[-\frac{i\lambda}{2}\sum_{i,\ell}\Phi_{i,\ell}\right]
\det\left[\mathbb{1}+\prod_{\ell=L_T}^{1} e^{i\lambda\Phi_{\ell}}e^{-\Delta\tau K}\right] ,
\label{eq:app_hubbard_wup}
\end{equation}
where $K$ is the single-particle hopping matrix and the descending product $\prod_{\ell=L_T}^{1}$ follows the convention $B_{L_T}\cdots B_1$ of Sec.~\ref{sec:overview-dqmc}; the circuit construction in Appendix~\ref{app:state_prep_oracle} uses the opposite vertex-kinetic ordering, which differs by a similarity transformation and leaves the determinant unchanged. Two prefactors must be distinguished. The configuration-\emph{independent} factor $(e^{\Delta\tau U/4}/2)^{N_sL_T}$ is immaterial, since only ratios between configurations matter. The phase $\exp[-i\lambda\sum_{i,\ell}\Phi_{i,\ell}/2]$, by contrast, is configuration-\emph{dependent} and must be retained: it is the contribution of the $-1/2$ shift, and it is precisely what renders the two flavor weights complex conjugates below, cancelling in their product.
 
At half filling on a bipartite lattice $\Lambda=A\cup B$, with every nearest-neighbor bond connecting $A$ and $B$, we employ the staggered particle-hole transformation
\begin{equation}
P^{-1}c_{i\sigma}P=\eta_i\,c^{\dagger}_{i\sigma},
\qquad
\eta_i=
\begin{cases}
+1, & i\in A,\\
-1, & i\in B,
\end{cases}
\label{eq:app_hubbard_PH}
\end{equation}
so that $\eta_i\eta_j=-1$ on nearest-neighbor bonds. Combined with the spin-$1/2$ time reversal
\begin{equation}
\begin{cases}
T^{-1}c_{i\uparrow}T&=c_{i\downarrow} \\
T^{-1}c_{i\downarrow}T&=-c_{i\uparrow}
\end{cases}
,~~~
T\,i\,T^{-1}=-i,
\label{eq:app_hubbard_TRS}
\end{equation}
the composite $\Theta=PT$ is anti-unitary with $\Theta^{2}=-1$; the Kramers structure guarantees that the two flavor sectors are genuinely paired rather than trivially relabelled, which is what makes the factorization of Eq.~\eqref{eq:app_hubbard_positive_weight} below nontrivial. The essential action of $\Theta$ is to flip both the imaginary unit and the centered density: $\Theta^{-1}i\,\Theta=-i$ and $\Theta^{-1}\delta n_{i\uparrow}\Theta=-\delta n_{i\downarrow}$.
 
Both the hopping term and the density vertex are invariant under $\Theta$. For the hopping, $P$ maps $c^{\dagger}_{i\sigma}c_{j\sigma}\mapsto \eta_i\eta_j\, c_{i\sigma}c^{\dagger}_{j\sigma}=-\eta_i\eta_j\,c^{\dagger}_{j\sigma}c_{i\sigma}$, and since $\eta_i\eta_j=-1$ on every bond this returns $c^{\dagger}_{j\sigma}c_{i\sigma}$; summing over bonds and using the Hermiticity of $K$ leaves the kinetic term unchanged, while $T$ merely relabels the spin sum. For the vertex, $\Theta^{-1}\big[i\lambda\Phi_{i,\ell}\,\delta n_{i\uparrow}\big]\Theta = (-i)\lambda\Phi_{i,\ell}(-\delta n_{i\downarrow}) = i\lambda\Phi_{i,\ell}\,\delta n_{i\downarrow}$, i.e. the exponent maps into its spin-flipped counterpart with the same sign, so the product over both flavors in Eq.~\eqref{eq:app_hubbard_hs_identity} is invariant. This HS decomposition is therefore sign-problem-free, in the symmetry class of Refs.~\cite{WuZhang2005SignFree,LiJiangYao2015Majorana,Wei2016MajoranaPositivity}.
 
More importantly, $\Theta$ maps the spin-up block of the action matrix to the complex conjugate of the spin-down block, so that the spin-up weight of Eq.~\eqref{eq:app_hubbard_wup} is the complex conjugate of the spin-down weight,
\begin{equation}
 w_\downarrow[\Phi]=w_\uparrow[\Phi]^*,
 \qquad
 W_F[\Phi]=w_\uparrow[\Phi]w_\downarrow[\Phi]=|w_\uparrow[\Phi]|^2 .
\label{eq:app_hubbard_positive_weight}
\end{equation}
This is the concrete realization of the squared-modulus structure exploited in the main text: in the DQ$^{2}$MC protocol only the spin-up block of the action matrix need be computed, and the Born rule supplies the second flavor.
 
The same symmetry reconstructs spin-down Green's functions from the spin-up block, from which correlation functions of the form of Eq.~\eqref{eq:def_corr} follow by Wick's theorem. Denoting the imaginary-time Green's function for spin $\sigma$ by,
\begin{equation}
G^\sigma_{ij}(\tau;\Phi)=\langle c_{i\sigma}(\tau)c^{\dagger}_{j\sigma}(0)\rangle_\Phi ,
\end{equation}
one finds, writing the staggered sign explicitly in indices, for $0<\tau<\beta$,
\begin{equation}
G^\downarrow_{ij}(\tau;\Phi)=\eta_i\,\eta_j\left[G^\uparrow_{ji}(\beta-\tau;\Phi)\right]^*,
\label{eq:app_hubbard_Gdown_tau}
\end{equation}
while the equal-time discontinuity gives
\begin{equation}
G^\downarrow_{ij}(0^+;\Phi)=\delta_{ij}-\eta_i\,\eta_j\left[G^\uparrow_{ji}(0^+;\Phi)\right]^* .
\label{eq:app_hubbard_Gdown_equal_time}
\end{equation}
Equivalently, with $\eta_i=e^{i\mathbf{Q}\cdot\mathbf{r}_i}$ and $\mathbf{Q}=(\pi,\pi)$,
\begin{equation}
G^\downarrow(\mathbf{q},i\omega_n;\Phi)
=-\left[G^\uparrow(\mathbf{q}-\mathbf{Q},i\omega_n;\Phi)\right]^* ,
\label{eq:app_hubbard_Gdown_qw}
\end{equation}
where $\omega_n=(2n+1)\pi/\beta$ is a fermionic Matsubara frequency; the minus sign is only the fermionic phase $e^{i\omega_n\beta}=-1$. Bosonic frequencies $\Omega_n=2\pi n/\beta$ will appear below for the two-particle correlators.
 
For the connected density-density correlator the $-1/2$ shift drops out after taking the connected part. At fixed HS configuration, Wick contraction gives the same-spin correlator
\begin{equation}
N^{\mathrm{conn}}_{\sigma\sigma}(i,j;\tau|\Phi)
=G^\sigma_{ji}(\beta-\tau;\Phi)G^\sigma_{ij}(\tau;\Phi),
~~ 0<\tau<\beta .
\label{eq:app_hubbard_density_wick_real}
\end{equation}
Using Eq.~\eqref{eq:app_hubbard_Gdown_tau}, the staggered signs cancel between the two Green's functions, so that
\begin{equation}
N^{\mathrm{conn}}_{\downarrow\downarrow}(\mathbf{q},i\Omega_n|\Phi)
=
\left[N^{\mathrm{conn}}_{\uparrow\uparrow}(\mathbf{q},i\Omega_n|\Phi)\right]^* ,
\label{eq:app_hubbard_Ndown_bubble}
\end{equation}
and the same-spin contribution to the charge channel is
\begin{equation}
N^{\mathrm{conn}}_{\uparrow\uparrow}(\mathbf{q},i\Omega_n|\Phi)
+
N^{\mathrm{conn}}_{\downarrow\downarrow}(\mathbf{q},i\Omega_n|\Phi)
=
2\Re N^{\mathrm{conn}}_{\uparrow\uparrow}(\mathbf{q},i\Omega_n|\Phi).
\label{eq:app_hubbard_charge_bubble_reconstruct}
\end{equation}
Furthermore, after averaging over $\Phi$ the real-space bubble is real and possesses both the spatial parity $\mathbf{r}\leftrightarrow-\mathbf{r}$ (square-lattice inversion) and the time symmetry $\tau\leftrightarrow\beta-\tau$ inherited from the swap symmetry $N^{\mathrm{conn}}_{\sigma\sigma}(i,j;\tau|\Phi)=N^{\mathrm{conn}}_{\sigma\sigma}(j,i;\beta-\tau|\Phi)$ of Eq.~\eqref{eq:app_hubbard_density_wick_real}. Hence $\sum_{\Phi} N^{\mathrm{conn}}_{\uparrow\uparrow}(\mathbf{q},i\Omega_n|\Phi)$ is real, and the charge correlator is twice the spin-up correlator. The single-particle Green's function carries the staggered momentum shift $\mathbf{Q}$, but the density one does not, since the density is local and the staggered factors $\eta_i$ appear twice inside the Wick contraction. Other correlators are reconstructed in similar fashion.

\subsection{Benchmark Model and Parameters}\label{app:hubbard_setup}

The simulation is performed at the state-vector level, where the squared modulus of each component gives the probability of the corresponding auxiliary-field configuration.
We stress that throughout this work, the benchmark results purely serve as proof-of-demonstration for the DQ$^2$MC algorithm. The bottlenecks in the tested system size are entirely limited by the resources in classical simulation; the scaling on quantum hardware is addressed in Secs.~\ref{sec:qalg_summary} and~\ref{sec:compat}.

\begin{figure}[t]
\centering
\includegraphics[width=7cm]{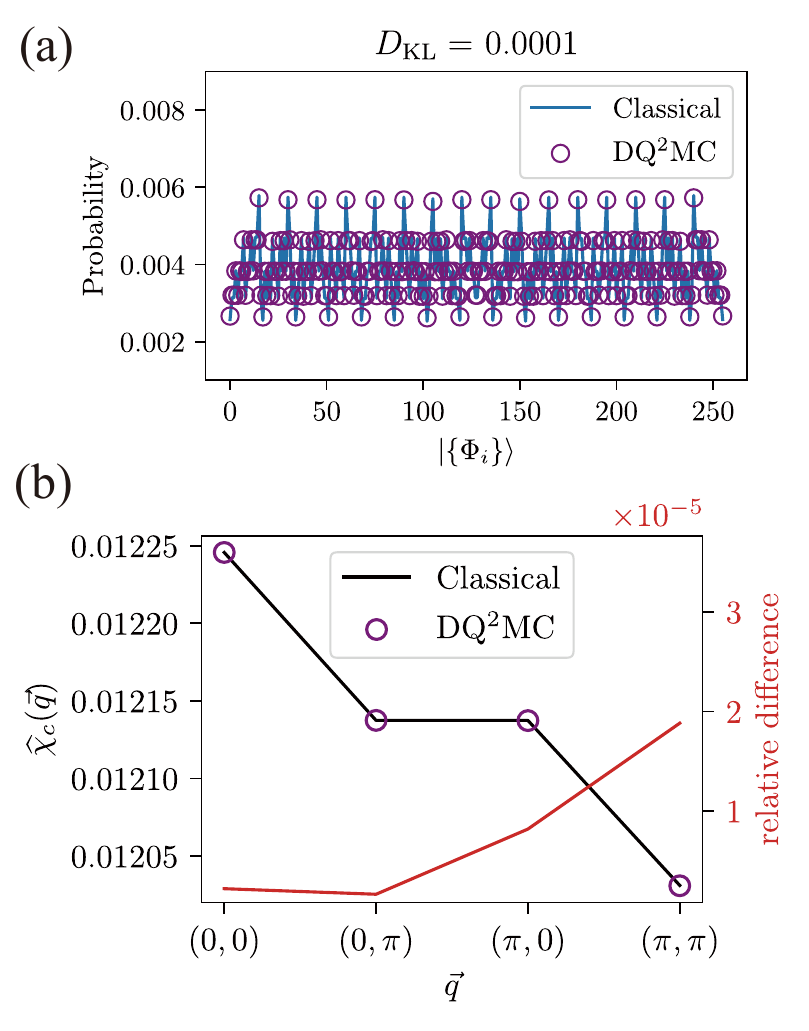}
\caption{\label{Fig::fig_fullq_bench} {\bf The Hubbard benchmark using the full quantum protocol.} (a) Prepared DQ$^{2}$MC probability distribution versus the exact classical DQMC weights over all $256$ HS configurations; the Kullback--Leibler divergence is $D_{\mathrm{KL}}=10^{-4}$. (b) Density correlation estimator $\widehat{\chi}_{c}(\q)$ [Eq.~\eqref{eq:suppl_hada_readout}] at the four high-symmetry momenta of the 2D Brillouin zone, defined in Eq.~\eqref{eq:suppl_hada_readout}; the Hadamard-test readout matches the direct classical evaluation to a relative error $\sim 10^{-5}$ (right axis).}
\end{figure}

We show the benchmark result for the full quantum protocol in Fig.~\ref{Fig::fig_fullq_bench}.
Panel (a) compares the resulting Born distribution with exact classical DQMC over all $256$ auxiliary-field configurations. Defining
$D_{\mathrm{KL}}(p\|p^{\mathrm{prep}})\equiv\sum_{\Phi}p_{\Phi}\log[p_{\Phi}/p_{\Phi}^{\mathrm{prep}}]$, with $p$ the exact distribution and $p^{\mathrm{prep}}$ the prepared distribution, gives $D_{\mathrm{KL}}=10^{-4}$. Since the calculation is a noiseless state-vector emulation, the residual discrepancy reflects only the finite-QSVT approximation.
In Fig.~\ref{Fig::fig_fullq_bench}(b), we benchmark Eq.~\eqref{eq:qalg_hadamard} on the $2\times 2$ repulsive Hubbard system, with $O$ the density bilinear and $A_{\Phi}=B_{\Phi}=G_{\Phi}\equiv M_{\Phi}^{-1}$, so that $\overline{O}(\q)$ is the connected density correlation $\chi_{c}(\q)$. The plotted readout is the Hadamard test estimator $\widehat{\chi}_c(\q)$, defined as
\be
\label{eq:suppl_hada_readout}
\widehat{\chi}_c(\q)\equiv2p(0_{\mathrm{H}})-1=\mathrm{Re}[\overline{O}(\q)]/(\alpha_A\alpha_B)
\ee
, where $\alpha_A=\alpha_B=1/(C\,\alpha_M)$ and $C$ is the QSVT-inverse prefactor. It matches the direct classical evaluation at the four high-symmetry momenta of the 2D Brillouin zone to relative error $\sim 10^{-5}$.

\begin{figure}[t]
\centering
\includegraphics[width=7cm]{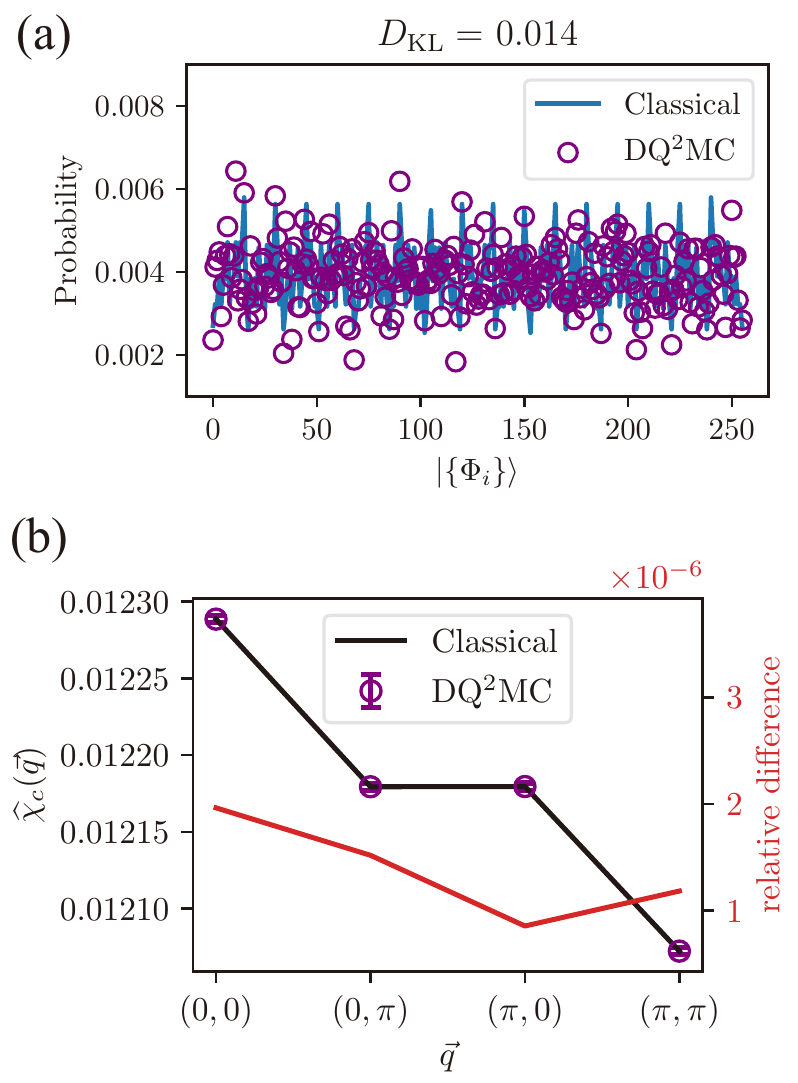}
\caption{\label{Fig::fig_hyb_bench} {\bf The Hubbard benchmark using the hybrid protocol.} (a) Marginal HS probability distribution from the hybrid chain versus the classical DQMC weights over all $256$ configurations; the Kullback--Leibler divergence is $D_{\mathrm{KL}}=1.4\times 10^{-2}$. (b) Density correlation estimator $\widehat{\chi}_{c}(\q)$ at the four high-symmetry momenta of the 2D Brillouin zone [Eq.~\eqref{eq:suppl_hada_readout}]; the hybrid readout matches the direct classical evaluation to a relative error $\sim 10^{-6}$ (right axis).}
\end{figure}

We show the benchmark on the same $2\times 2$ Hubbard system for the hybrid protocol in Fig.~\ref{Fig::fig_hyb_bench}. In panel (a), we compare the cluster-Gibbs histogram with the exact DQMC weights and obtain $D_{\mathrm{KL}}=1.4\times10^{-2}$. In panel (b), we benchmark $\widehat{\chi}_c(\q)$ obtained from the Hadamard-test estimator defined in Eq.~\eqref{eq:compat_hyb_obs} with classical averaging for the density correlation at the four high-symmetry momenta; the relative error is $\sim10^{-6}$.

\subsection{Preconditioning and Postselection}\label{app:hubbard_preconditioning}

Here we detail how the rescaling parameter $x_0$ of Eq.~\eqref{eq:qalg_x0} and the corresponding $\alpha_{\mathrm{rescale}}$ is tuned to optimize the state-preparation success probability without degrading the fidelity of the resulting state. Here, $x_0>0$ globally rescales the diagonal block encoding of $p_{\Phi}$ in $\mathcal{O}_{\det}$, as schematically showed in Fig.~\ref{Fig::precon} in the main text.

\begin{figure}[t]
\centering
\includegraphics[width=8cm]{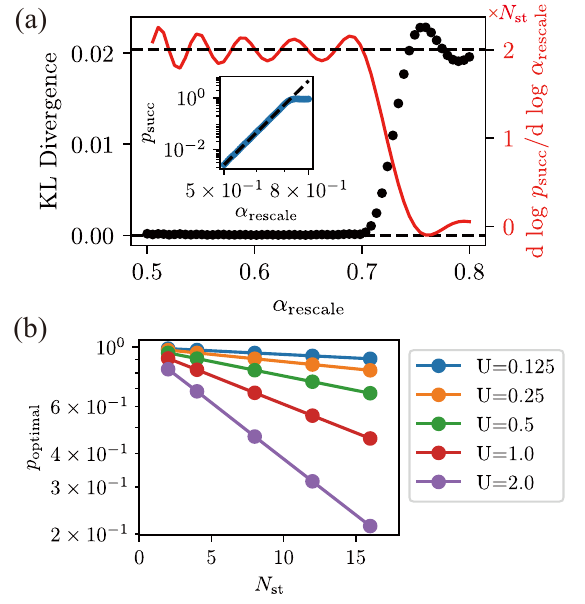}
\caption{\label{Fig::fullq_o_op} {\bf Preconditioning of the determinant-oracle state preparation.} (a) KL divergence between the prepared and exact DQMC distributions (black) and the logarithmic derivative $\mathrm{d}\log p_{\mathrm{succ}}/\mathrm{d}\log\alpha_{\mathrm{rescale}}$ (red, in units of $N_{\tn{st}}$). In the faithful regime the derivative sits around $2N_{\tn{st}}$ and the KL divergence is negligible; the derivative drops and the KL divergence rises as the window begins to distort the high-weight tail. Inset: log-log $p_{\mathrm{succ}}$ showing the power-law-to-plateau crossover. (b) $p_{\mathrm{optimal}}$ as a function of $N_{\tn{st}}$ for several interaction strengths $U$ ; the exponential decay steepens with $U$, suggesting the increasing condition number of the DQMC distribution.}
\end{figure}

Practically, the exp-QSVT filter $f_{\exp}(x)\simeq e^{N_{\tn{st}}(x+x_0)}$ is implemented as a windowed approximation (see Appendix~\ref{app:state_prep_oracle}) clipped to unity for $x>-x_0$. Thus, the optimal choice for $x_0$ is reached when $\max_{\Phi} f_{\exp}(x_{\Phi})=1$, at which point the success probability is denoted $p_{\tn{optimal}}$.
As long as $\alpha_{\mathrm{rescale}}$ is chosen below the threshold set by $\max_{\Phi}x_{\Phi}<-x_0$, the window is inactive and the filter acts faithfully on the full range of determinant weights.
This clipping also gives a practical route to the optimal $\alpha_{\tn{rescale}}$. Increasing $\alpha_{\tn{rescale}}$ from a small value does not change the fidelity of the prepared state, as seen from the negligible KL divergence in Fig.~\ref{Fig::fullq_o_op}(a). Above a threshold in $\alpha_{\tn{rescale}}$ the KL divergence rises sharply; this threshold marks the optimum. The optimum can be pinpointed by tracking the scaling of $p_{\tn{succ}}$: below the optimum, $p_{\tn{succ}}\propto\alpha_{\tn{rescale}}^{2N_{\tn{st}}}$ from Eq.~\eqref{eq::det_oracle_action} [Fig.~\ref{Fig::fullq_o_op}(b) inset], whereas above it the window clips the highest-weight configurations and the scaling exponent deviates from $2N_{\tn{st}}$ [red curve, Fig.~\ref{Fig::fullq_o_op}(b)]. The onset of this deviation identifies the optimal rescaling factor; in practice, a few trial runs scanning $\alpha_{\tn{rescale}}$ and tracking $p_{\tn{succ}}$ suffice to estimate it before a production run.

We now derive the exact optimal success probability quoted in Eq.~\eqref{eq:qalg_psucc}. From Eq.~\eqref{eq::det_oracle_action}, the rescaled diagonal block-encoding weight is
\begin{equation}
d_{\Phi}\equiv \alpha^{2N_{\tn{st}}}_{\tn{rescale}} |\det M_{\Phi}|^{2}\leq 1,
\qquad
d_{\Phi}\propto p_{\Phi},
\label{eq:qalg_dphi}
\end{equation}
so that the success probability of Eq.~\eqref{eq::det_oracle_action} is
\begin{equation}
p_{\mathrm{succ}}
=
\frac{1}{2^{N_{\tn{st}}}}\sum_{\Phi}d_{\Phi} .
\label{eq:qalg_psucc_raw}
\end{equation}
At the optimum, $\alpha_{\tn{rescale}}$ is fixed by saturating the largest weight, $d_{\max}\equiv\max_{\Phi}d_{\Phi}=1$, i.e. $\alpha_{\tn{rescale}}^{2N_{\tn{st}}}=1/|\det M|^{2}_{\max}$, which makes $d_{\Phi}=p_{\Phi}/p_{\max}$ exactly. Since $\sum_{\Phi}p_{\Phi}=1$, Eq.~\eqref{eq:qalg_psucc_raw} gives Eq.~\eqref{eq:qalg_psucc}. The last inequality there, obtained by replacing every $p_{\Phi}$ with $p_{\min}$, recovers the looser condition-number bound.

The extreme case $p_{\max}=2^{-N_{\tn{st}}}$ (a uniform distribution, achievable at $\alpha_{\tn{rescale}}=1$) gives $p_{\mathrm{succ}}=1$, independent of $N_{\tn{st}}$. However, if the distribution is concentrated at several configurations, then $p_{\mathrm{succ}}$ can be exponentially small at worst. The benchmark is an example of this unfavorable regime. The exponential decay of $p_{\mathrm{succ}}$ in Fig.~\ref{Fig::fullq_o_op}(b) indicates, by Eq.~\eqref{eq:qalg_psucc}, that $p_{\max}/2^{-N_{\tn{st}}}$ grows exponentially with the space-time volume. Increasing $U$ raises $p_{\max}$ and therefore steepens this decay, as observed across the plotted curves.

\section{Degree Scaling for the Exp-QSVT Polynomial}\label{app:exp_qsvt_degree}
 
We bound the polynomial degree, and hence the QSVT query depth, needed to approximate the exponential window of Eq.~\eqref{eq:qalg_expwindow} on a bounded interval. Writing the target in the generic form $f(s)=e^{bs-a}$ for $s\in[0,s_{\max}]$, the dictionary to the loading construction of Sec.~\ref{sec:loading} is
\begin{equation}
s \leftrightarrow x_{\Phi},
\qquad
b = N_{\tn{st}}\,\alpha_{\log},
\qquad
a = -N_{\tn{st}}\,x_{0},
\label{eq:app_dictionary}
\end{equation}
where the factor $\alpha_{\log}$ accounts for the fact that the exp-QSVT acts on the rescaled variable $x_{\Phi}/\alpha_{\log}$ defined in Appendix~\ref{app:state_prep_oracle}. Since $\alpha_{\log}$ is an intrinsic $O(1)$ constant, it does not affect the scaling with $N_{\tn{st}}$ below. Throughout this appendix $\varepsilon\equiv\varepsilon_{\exp}$ denotes the exp-QSVT tolerance, distinct from the log-QSVT tolerance $\varepsilon_{\log}$ defined in Appendix~\ref{app:state_prep_oracle}. We work in the regime $bs_{\max}-a\le 0$, so that $|f(s)|\le 1$ on the interval and no clipping is required; as discussed at the end of this appendix, this is precisely the faithful operating regime identified in Sec.~\ref{sec:loading}.

Let us start by rescaling the interval.
Define $t\equiv s/s_{\max}\in[0,1]$, so that $s=s_{\max}t$ and
\begin{equation}
f(s)=e^{bs_{\max}t-a}=e^{bs_{\max}-a}\,e^{-\nu(1-t)},
\qquad
\nu\equiv bs_{\max},
\end{equation}
where we write $\nu$ for the total exponent variation across the interval, reserving $\beta$ for the inverse temperature. The nontrivial approximation problem is therefore the shifted exponential
\begin{equation}
F(t)\equiv e^{-\nu(1-t)},
\qquad
t\in[0,1].
\end{equation}

Next, let us bound the degree of polynomial that is required to approximate $F(t)$. 
The standard QSVT-compatible exponential approximation~\cite{GilyenSuLowWiebe2019QSVT, LowChuang2017QSP} provides a polynomial $P(t)$ with
\begin{equation}
\sup_{t\in[0,1]}|P(t)-e^{-\nu(1-t)}|\le \varepsilon' ,
\end{equation}
of degree
\begin{equation}
\deg(P)
=
O\!\left(\sqrt{\max\{\nu,\log(1/\varepsilon')\}\,\log(1/\varepsilon')}\right).
\label{eq:app_degP}
\end{equation}

With the degree scaling in hand, let us multiply it by the prefactor and restore the variable $s$.
Define the lifted approximant
\begin{equation}
\widetilde f(s)\equiv e^{bs_{\max}-a}\,P\!\left(\tfrac{s}{s_{\max}}\right),
\end{equation}
for which
\begin{equation}
\sup_{s\in[0,s_{\max}]}|f(s)-\widetilde f(s)|
\le e^{bs_{\max}-a}\,\varepsilon' .
\end{equation}
Achieving absolute error $\varepsilon$ therefore requires $\varepsilon'\le \varepsilon\,e^{a-bs_{\max}}$. In the bounded regime $bs_{\max}-a\le 0$ the prefactor satisfies $e^{bs_{\max}-a}\le 1$, so choosing $\varepsilon'=\varepsilon$ already suffices, and the exp-QSVT degree $d_{\exp}$ --- equivalently the QSVT query depth --- obeys
\begin{equation}
d_{\exp}
=
O\!\left(\sqrt{\max\{bs_{\max},\log(1/\varepsilon)\}\,\log(1/\varepsilon)}\right).
\label{eq:app_DQSVT_general}
\end{equation}
 
The natural choice for the loading construction of Sec.~\ref{sec:loading} is the saturation cutoff $s_{\max}=a/b$, at which $f(s_{\max})=1$ exactly --- the condition $\max_{\Phi}f_{\exp}(x_{\Phi})=1$ defining the optimal preconditioning. Substituting $bs_{\max}=a$ into Eq.~\eqref{eq:app_DQSVT_general},
\begin{equation}
d_{\exp}
=
O\!\left(\sqrt{\max\{a,\log(1/\varepsilon)\}\,\log(1/\varepsilon)}\right).
\label{eq:app_DQSVT_x0}
\end{equation}
For $a\gtrsim\log(1/\varepsilon)$ this reduces to $d_{\exp}=O(\sqrt{a\,\log(1/\varepsilon)})$, and for $a\lesssim\log(1/\varepsilon)$ to $d_{\exp}=O(\log(1/\varepsilon))$.
 
The controlling parameter is thus not $b$ alone but the total exponent variation $\nu=bs_{\max}$ across the interval, which equals $a$ at the saturation cutoff. From the dictionary of Eq.~\eqref{eq:app_dictionary}, both $a$ and $b$ are extensive --- proportional to $N_{\tn{st}}$, with $\alpha_{\log}$ and $x_0$ intrinsic --- while their ratio $a/b=-x_0$ is intensive. Hence
\begin{equation}
d_{\exp}=O\!\left(\sqrt{N_{\tn{st}}\,\log(1/\varepsilon_{\exp})}\right),
\end{equation}
which is the $\kappa_{\exp}\sim\sqrt{N_{\tn{st}}}$ factor quoted in Sec.~\ref{sec:qalg_summary} and the origin of the $N_{\tn{st}}^{1/2}$ enhancement over the naive gate count.
 
Finally, the analysis above assumes $bs_{\max}-a\le 0$, i.e. that the exponential window does not exceed unity anywhere on the interval and no clipping is active. By Eq.~\eqref{eq:app_dictionary} this is exactly the condition $\max_{\Phi}x_{\Phi}\le -x_{0}$ identified in Sec.~\ref{sec:loading} as the faithful regime, in which the filter acts without distorting the high-weight tail --- and which the preconditioning protocol there is designed to locate. Above the optimum the implemented window is clipped to unity and the present bound no longer applies.

\section{Path integral formulation of PQMC}\label{app:pqmc_path_integral}
 
We outline the coherent-state path-integral derivation that produces the extended action matrix $M_{\Phi}^{(0)}$ of Eq.~\eqref{eq:outlook_M_pqmc_defM} from the Slater-determinant PQMC weight. After Trotterizing the imaginary-time evolution and HS-decoupling the interaction at each slice, the projector weight reads
\begin{equation}
p_{\Phi}\propto\langle L|\,e^{-\Delta\tau H(\Phi_{L_{T}})}\cdots e^{-\Delta\tau H(\Phi_{1})}\,|R\rangle,
\label{eq:app_pqmc_trotter}
\end{equation}
with $|R\rangle\equiv\prod_{n=1}^{N_{p}}R_{nj}c^{\dagger}_{j}|0\rangle$ and $|L\rangle\equiv\prod_{n=1}^{N_{p}}L_{nj}c^{\dagger}_{j}|0\rangle$ the trial Slater determinants. Inserting Grassmann coherent-state resolutions of identity,
\begin{equation}
\mathbb{1}=\int d\bar{\xi}\,d\xi\;e^{-\bar{\xi}\xi}\,|\xi\rangle\langle\xi|,
\end{equation}
between every pair of slice operators recasts Eq.~\eqref{eq:app_pqmc_trotter} as a Grassmann path integral over $\{\bar{\xi}_{\tau},\xi_{\tau}\}_{\tau=0}^{L_{T}}$,
\begin{widetext}
\begin{equation}
p_{\Phi}\propto
\int\!\prod_{\tau=0}^{L_{T}}d\bar{\xi}_{\tau}\,d\xi_{\tau}\;
e^{-\sum_{\tau}\bar{\xi}_{\tau}\xi_{\tau}}\;
\langle L|\xi_{L_{T}}\rangle\,
\Big[\prod_{\tau=1}^{L_{T}}\langle\xi_{\tau}|e^{-\Delta\tau H(\Phi_{\tau})}|\xi_{\tau-1}\rangle\Big]\;
\langle\xi_{0}|R\rangle.
\label{eq:app_pqmc_path}
\end{equation}
\end{widetext}
Unlike the finite-temperature path integral of Sec.~\ref{sec:overview-dqmc}, the temporal index here runs over an open strip rather than a torus, with two boundary overlaps $\langle L|\xi_{L_{T}}\rangle$ and $\langle\xi_{0}|R\rangle$.
 
Each boundary overlap with a Slater-determinant trial reduces to a product of $\delta$-functions via standard Grassmannian integration rules, enforcing that $\xi$ lies in the subspace spanned by the trial orbitals. This $\delta$-function boundary condition can be obtained by introducing two auxiliary Grassmann vectors 
\begin{equation}
\langle L|\xi_{L_{T}}\rangle=\int d\bar{\chi}\;e^{-\bar{\chi}L^{\dagger}\xi_{L_{T}}},
~~
\langle\xi_{0}|R\rangle=\int d\chi\;e^{-\bar{\xi}_{0}R\chi},
\label{eq:app_pqmc_proj}
\end{equation}
up to a $\Phi$-independent sign fixed by the ordering convention of the Grassmann measure, which is immaterial since only ratios of weights between configurations matter. The effect of Eq.~\eqref{eq:app_pqmc_proj} is to trade the two boundary terms for two additional Gaussian integration variables per filled orbital in the trial wavefunction, at the cost of bordering the action matrix.
 
Substituting Eq.~\eqref{eq:app_pqmc_proj} into Eq.~\eqref{eq:app_pqmc_path} and using the coherent-state matrix element $\langle\xi_{\tau}|e^{-\Delta\tau H(\Phi_{\tau})}|\xi_{\tau-1}\rangle\propto\exp(\bar{\xi}_{\tau}B_{\tau}\xi_{\tau-1})$, with $B_{\tau}$ the single-slice propagator of Sec.~\ref{sec:overview-dqmc}, in the same vertex-then-kinetic ordering used there, and with the suppressed coherent-state normalizations independent of $\Phi$ produces a Gaussian Grassmann integral with action $\bar{\Xi}\,M_{\Phi}^{(0)}\,\Xi$ over the augmented vector
\begin{equation}
\Xi\equiv(\xi_{0},\xi_{1},\ldots,\xi_{L_{T}},\chi)^{\mathsf{T}},
~~
\bar{\Xi}\equiv(\bar{\xi}_{0},\bar{\xi}_{1},\ldots,\bar{\xi}_{L_{T}},\bar{\chi}),
\end{equation}
with $M_{\Phi}^{(0)}$ exactly the matrix of Eq.~\eqref{eq:outlook_M_pqmc_defM}. Since $\Xi$ carries $N_{\tn{st}}$ field components together with the $N_{p}$ boundary components, $M_{\Phi}^{(0)}$ is square of dimension $(N_{\tn{st}}+N_{p})$, as used in Sec.~\ref{sec:pq2mc}.
 
Carrying out the Gaussian Grassmann integral gives $p_{\Phi}\propto\det M_{\Phi}^{(0)}$, all suppressed prefactors being configuration-independent. Taking the Schur complement on the $\xi$-block --- that is, integrating out the $N_{\tn{st}}$ bulk components --- collapses the bordered matrix to an $N_{p}\times N_{p}$ determinant and recovers the closed-form weight $p_{\Phi}\propto \det[L^{\dagger}B_{\Phi}(2\Theta,0)R]$ of Eq.~\eqref{eq:outlook_M_pqmc}. The same construction maps the time-displaced Green's function to ${[M_{\Phi}^{(0)}]}^{-1}$ block by block, yielding Eq.~\eqref{eq:outlook_G_pqmc} upon inverting the bordered block matrix.

\bibliographystyle{apsrev4-1_custom}
\bibliography{refs}
\clearpage

\end{document}


\renewcommand{\thefigure}{S\arabic{figure}}
\renewcommand{\figurename}{Supplemental Figure}
\setcounter{figure}{0}
\setcounter{section}{0}
\newcounter{suppfigure} 
\renewcommand{\thesuppfigure}{S\arabic{suppfigure}} 
\begin{widetext}
\begin{center}
    {\bf Supplementary material for ``Theory of Correlated Insulators and Superconductor at $\nu=1$ in Twisted WSe$_2$"}\\
    Sunghoon Kim$^*$, Juan Felipe Mendez-Valderrama$^*$,  Xuepeng Wang$^*$, Debanjan Chowdhury
\end{center}

\section{Details of parton mean-field calculations}
\label{sec:SI_parton}

\end{widetext}

\bibliographystyle{apsrev4-1_custom}
\bibliography{refs}